\documentstyle[preprint,aps,tighten,epsf,floats]{revtex}

\begin{document}

\thispagestyle{empty}

\vspace{2cm}

\title{The $N N \rightarrow NN \pi^+ $ Reaction near Threshold in 
a Chiral Power Counting Approach}

\author{Carlos A. da Rocha$^a$,  
Gerald A. Miller$^b$, and
Ubirajara van Kolck$^{c,b}$ }

\vspace{1cm}

\address{$^a$ Instituto de F\'{\i}sica Te\'orica, Universidade Estadual
Paulista \\ Rua Pamplona 145, 01405-900 S\~ao Paulo-SP, Brazil \\ and \\
Departamento de Desenvolvimento Tecnol\'ogico, Universidade S\~ao Judas Tadeu\\
Rua Taquari 546, 03166-000 S\~ao Paulo-SP, Brazil}

\vspace{0.3cm}
\address{$^b$ Department of Physics \\ University of 
Washington, Box 351560, Seattle, WA 98195-1560}

\vspace{0.3cm}

\address{$^c$ Kellogg Radiation Laboratory, 106-38 \\
California Institute of Technology, Pasadena, CA 91125}

\maketitle

\vspace{2cm}

\begin{abstract} 
Power-counting arguments are used to organize the interactions
contributing to the 
$N N \rightarrow d \pi, pn \pi$ reactions near threshold. 
We estimate the contributions
from the
three formally leading mechanisms: 
the Weinberg-Tomozawa (WT) term, the impulse term, 
and the $\Delta$-excitation mechanism. 
Sub-leading but potentially 
large mechanisms, including $S$-wave pion-rescattering, 
the Galilean correction to the 
WT term, and short-ranged contributions are also examined.
 The  WT term is shown to be numerically the largest,
and the other contributions are found to approximately cancel. 
Similarly to the reaction
$pp \rightarrow pp\pi^0$, the computed 
cross sections are considerably smaller than  the data.
We discuss possible origins of this discrepancy.
\end{abstract}

\vspace{2cm}

\hfill{NT@UW-98-10}

\hfill{KRL MAP-236}

\hfill{IFT - P.058/98}

\newpage

\section{Introduction}

Computations of pion production in nucleon-nucleon 
($NN$) collisions near threshold, allow a  confrontation of 
our understanding of $NN$ interactions with data
in a kinematic region for which
chiral symmetry, and therefore quantum chromodynamics
QCD,  could be very important.
The reaction 
$pp \rightarrow pp\pi^0$ near threshold has attracted much
attention in the years since the first IUCF data appeared
\cite{M90} and has exposed  serious disagreements  with earlier 
theoretical calculations \cite{KR66,SSY69,MS91,N92}. 
The existence of many conflicting models claiming to explain 
this discrepancy \cite{LR92,HGM94,HO95,Han+95} 
calls for a principle to organize the several 
potentially significant mechanisms of pion production.
Chiral Perturbation Theory ($\chi$PT) has been applied to
mesonic \cite{W79,GL84}, one-baryon \cite{JM91,BKM95}, and 
nuclear \cite{W90,OK92,ORV94,V94,vKolckn,KSW,SKS98,monster}
processes where typical momenta of the order of the pion mass, $m_\pi$,
allow a systematic expansion of observables in powers of $m_\pi/M_{QCD}$,
where $M_{QCD}\sim 1$ GeV. 
Cohen {\it et al} \cite{bira} have adapted the power counting 
and applied $\chi$PT to 
near-threshold pion production, where momenta are of order $\sqrt{m_\pi m_N}$,
$m_N$ being the nucleon mass.
They estimated leading and next-to-leading contributions,
the latter including important short-range contributions,
related to the isoscalar components of the
potential (and possibly described by 
$\sigma$ and $\omega$ meson exchanges) \cite{LR92}.
Subsequently, van Kolck {\it et al} \cite{KMR96}  showed that
next-to-next-to-leading contributions ({\it e.g.}, from $\rho-\omega$
meson exchange) are also relevant; 
data could then be explained within the very large theoretical uncertainties
associated with $S$-wave pion rescattering and the short-range structure of
the nuclear force.
Other $\chi$PT-inspired calculations 
have also stressed the importance of understanding
rescattering \cite{SC95}
and the effect of loops \cite{SC95,Han+98,moalem,cagadadoulf,newsc}.
Since these $\chi$PT-inspired calculations include
a larger number of contributions than other, model calculations,
one concludes that a  large theoretical uncertainty
plagues all calculations performed to date.

Given such uncertainties, it is natural to examine other channels using 
the same techniques. 
Here we are going to discuss $NN\rightarrow d\pi$ and $\rightarrow pn\pi$, 
which traditionally have been considered better understood
than $pp \rightarrow pp\pi^0$. 
We consider energies near
threshold where the calculation simplifies,
because the pion emerges mostly in an $S$ wave.  
We will show that an understanding of these channels
is still in the future.

We  adopt here the conventional nuclear approach
of grouping  all $NN$ interactions generated by mesons of 
small momentum in a
potential, 
while the contributions associated with energies comparable
to $m_\pi$ are accounted for in a kernel to be evaluated
between wavefunctions generated by the potential. 
Splitting the problem this way, one should still strive to 
calculate wavefunctions and kernels from the same theory or model,
otherwise ambiguities arise from off-shell extrapolations
(or equivalently, from nucleon-field redefinitions).
$\chi$PT is the only known tool for performing this task,
and at the same time is  consistent with QCD, because  
 its symmetries are treated correctly.
$\chi$PT is also unique in that it offers the possibility
of doing systematic calculations: an expansion in momenta
provides a power counting to organize the calculation even
though coupling constants are not small. 

$\chi$PT separates interactions in long-range effects
calculated explicitly with pion exchange 
and short-range effects accounted for by contact 
interactions with an increasing number of derivatives.
Parameters not constrained by chiral symmetry depend
on details of QCD dynamics, and are at the present
unknown functions of QCD parameters. Predictive power
is not lost, however, because at any given order 
in the power counting only 
a finite number of unknown parameters appear; after they 
are fitted to a finite set of data, all else can be predicted
at that order. These predictions are called 
``low-energy theorems''.
Since the Lagrangian of $\chi$PT is the most general
one consistent with QCD symmetries, $\chi$PT is a generalization
of current algebra.

Unfortunately, the only $NN$ potential derived
in $\chi$PT \cite{OK92} and fitted to
low-energy phase shifts \cite{ORV94} produces poor results
near the pion production threshold. 
Attempts to remedy the situation are in progress \cite{SKS98,monster}.
For the time being we will rely on modern, ``realistic''
phenomenological potentials which fit $NN$ data very well. 
By considering more than one of those, we can estimate
the otherwise uncontrolled error stemming
from our inconsistent use of potential and kernel.
We are going to see that, in contrast to neutral
pion production, the error is small in the channel considered here.
Because such realistic potentials reproduce low-energy phase shifts
with identical long-range tails, they must contain the equivalent
to the leading order in the chiral expansion. 
Although our approach should be considered phenomenological,
our leading-order result will be an approximation to a low-energy
theorem.
At sub-leading orders in the expansion, an apparent ambiguity arises,
concerning the correct treatment of the energy transferred
in pion rescattering. This
can be seen in the conflicting results of estimates of the same 
kernel
\cite{bira,SC95}. This issue is under study \cite{usandthem};
here we will limit ourselves to the most natural prescription
that the transferred energy is $\simeq m_\pi/2$.

We will concentrate most of our efforts on the kernel.
We will discuss a
reasonable power counting for pion production that
generalizes for $\pi d$ and $\pi pn$ channels the discussion of
Ref. \cite{bira}. In leading order, the new ingredient here is
pion rescattering using the Weinberg-Tomozawa (WT) 
term that dominates isospin-dependent $\pi N$ scattering. 
This term does not contribute to the reaction
$pp\rightarrow pp\pi^0$.
In the latter, the leading non-vanishing order consists
only of an impulse term (IA), in which a single pion 
is emitted from a nucleon, and a similar contribution
from the delta ($\Delta$).
This leading order underpredicts the 
experimental data by a factor of 
approximately 5, due to 
two cancellations not incorporated in our power counting:
{\it (i)} among different regions in coordinate space for each term
evaluated between initial and final wavefunctions;
and {\it (ii)} between the total impulse and delta contributions.
Oversight {\it (i)} results from our present inability
to treat the potential and the kernel on the same footing.
Oversight  {\it (ii)} is somewhat accidental, but actually expected 
from the fact that, in energy, the pion threshold sits
midway between the elastic threshold and the delta pole.
The sensitivity of the delta contribution to the realistic
potential used is the main source of dependence
on the $NN$ potential of the final result.  

As a consequence of the accidentally small leading order,
effects which are usually negligible acquire
prominence in neutral pion production in the $pp$ reaction.
One effect is isospin-independent
pion rescattering.
$\chi$PT is critically necessary to assess the size of this contribution.
First, in principle $\chi$PT allows one to determine from
$\pi N$ data not only the momentum- and energy-independent
$2\pi N^\dagger N$ vertex, but also terms which are quadratic in
energy and momentum. This is important in view
of the fact that in pion production
the virtual pion has energy of order $m_\pi/2$
and the combination of parameters
of relevance is thus different from the combination that
appears at the $\pi N$ scattering threshold.
There is by now a number of consistent
estimates of the relevant parameters to third order
\cite{BKM95,ulfandall}. An estimate
of the uncertainty of this contribution to pion production
can be made by using also a lower-order determination \cite{sub}.
Second, $\chi$PT is the only way to account for rescattering as a component 
of a Feynman diagram without destroying chiral symmetry. 
There have been attempts to estimate rescattering by
simply connecting a nucleon line to
a $\pi N$ amplitude with an arbitrary off-shell
extension \cite{HGM94,HO95}. By considering field redefinitions
in the most general chiral Lagrangian, it is easy to show
that the off-shell ambiguity in the pion leg is
equivalent to a set of short-range $\pi (N^\dagger N)^2$ interactions, 
and that an inconsistent treatment of both
effects leads to violation of chiral symmetry in a way
that is contradictory with QCD \cite{friar}. 
Using $\chi$PT, it was found \cite{bira,SC95} that rescattering interferes
destructively with the leading-order effects in 
$pp\rightarrow pp\pi^0$. This further 
interference makes agreement with data more difficult.
Although the magnitude of the effect
is still being assessed \cite{usandthem}, it is certain
that the uncertainty stemming from different determinations
of the $\chi$PT parameters is large \cite{KMR96}. 
 
This disagreement between theoretical evaluations and cross-section
data for $pp\rightarrow pp\pi^0$ can be largely removed
if $\sigma$, $\rho$ and other 
heavy mesons are included \cite{KMR96}.
When first suggested \cite{LR92},
it looked as if  $pp\rightarrow pp\pi^0$ was a clear signal 
of these otherwise elusive mechanisms.
Among the first corrections in our power counting,
one finds two-pion-exchange
loop graphs and  $\pi (N^\dagger N)^2$ counterterms (that behave
properly under chiral symmetry).
A full $\chi$PT calculation requires the calculation of these
loops, and although some steps have been taken in this 
direction \cite{moalem,newsc},
it is a herculean task that remains to be completed.
Even then, it will still require that the counterterms
be fitted to some pion production data (say, right at threshold)
so that other  data (say, the energy dependence close to threshold,
or other channels)
be predicted. 
In the case of $pp\rightarrow pp\pi^0$, even 
the most sophisticated phenomenological models 
have to recourse to 
such a fit of a short-range counterterm
\cite{hanhart}.
In any case, after all this one would then be interested
in determining whether such counterterms are of natural size,
and whether they can be further understood as the result of
heavy-meson exchange.
In view of all other uncertainties, the authors of Ref. \cite{bira}
took the point of view that 
an estimate of this class of sub-leading contributions
could more easily be made by modeling them
with meson exchange in Z-graphs, following Ref. \cite{LR92}.
It turns out that the counterterms so produced are of natural
size, which lends them credence, but that they are not 
sufficient to achieve agreement with data.
In order to study convergence, further meson exchanges
(chiefly $\rho-\omega$) that contribute to higher-order
counterterms were considered, and shown to be smaller but
still relevant \cite{KMR96}.

The conclusion of Ref.\cite{KMR96}
was that it is possible to describe the $pp\rightarrow pp\pi^0$ reaction
consistently with QCD, reasonable meson exchanges, and realistic
potentials, but only within a very large theoretical uncertainty.
It is our intention to assess here our theoretical understanding 
of the 
$N N \rightarrow d \pi, \rightarrow pn \pi$ reactions by 
analyzing the effect of the same microscopic mechanisms. 

These reactions have been studied for some time.
Conservation of parity, angular momentum, and isospin
constrains the possible channels for these reactions.
In the case of unbound final nucleons, sufficiently close
to threshold the strong $S$-wave two-nucleon interaction
implies that the most important channels should
be those in which the final nucleons
have relative orbital angular momentum $L_{NN}=0$.
Some of these partial waves are listed in Table~\ref{Tab.01}.

\begin{table}
\caption{Partial-wave amplitudes for the $NN\rightarrow NN\pi$ 
reactions
that can contribute to the isotropic cross section right at 
threshold with an $L_{NN}=0$ final state. 
$L_{\pi(NN)}$ denotes the pion angular momentum.}
\begin{tabular}{lccc}
Element & $NN$ Initial state & $NN$ final state & $L_{\pi(NN)}$ \\ \tableline
$a_0$   &         $^1S_0$    &      $^3S_1$     &       1      \\
$a_1$   &         $^3P_1$    &      $^3S_1$     &       0      \\
$a_2$   &         $^1D_2$    &      $^3S_1$     &       1      \\
$b_0$   &         $^3P_0$    &      $^1S_0$     &       0      
\end{tabular}
\label{Tab.01}
\end{table}

Near threshold we expect the pion, too, to be in an $S$-wave,
and the cross sections to be nearly isotropic, with \cite{MR55}
\begin{equation}
\frac{d\sigma}{d\Omega} (pp\rightarrow pn \pi^+)=
\frac{1}{4}\left(|a_1|^2+\frac{1}{3}|b_0|^2\right)
+\ldots,
\label{difcrosspn}
\end{equation}
\begin{equation}
\frac{d\sigma}{d\Omega} (pp\rightarrow d \pi^+)=
\frac{1}{4} |a_1'|^2+ \ldots,
\label{difcrossd}
\end{equation}
\begin{equation}
\frac{d\sigma}{d\Omega} (pp\rightarrow pp \pi^0)=
\frac{1}{12} |b_0|^2 +\ldots
\label{difcrosspp}
\end{equation}
Here $a_1$ and $a_1'$ have quite different magnitudes due to
the different wavefunctions; in the case of the deuteron final state
the effect of the $D$ wave is also important, and it is incorporated
in the following.
The cross section for $np\rightarrow d \pi^0$ is related to
that of $pp\rightarrow pn \pi^+$ by isospin considerations.
There is a relative factor
2 from $pp$ being pure
isospin $I=1$ and $np$ having $I=1$ and $I=0$ in equal
probabilities; and there are Coulomb effects.
The total cross-section for the deuteron channels is function of
$\eta=q/m_\pi$ where $q$ is the maximum pion momentum.
It is well-known \cite{GW54,R54} that sufficiently close to threshold
the cross-section for $n p \rightarrow d \pi^0 $ is
\begin{equation}
\sigma_{n p}= \frac{\alpha}{2}\eta + {\cal O}(\eta^3),
\label{snp}
\end{equation}
\noindent
while for $p p \rightarrow d \pi^+ $ it is
\begin{equation}
\sigma_{p p}= \alpha C_0^2 \eta + {\cal O}(\eta^3).
\label{spp}
\end{equation}
\noindent
Here $\alpha$ is a constant which depends on the strong
interactions that produce the pion
and $C_0(\eta)$ is 
a Coulomb factor arising mostly from the pion-deuteron electromagnetic
interaction. 
If the experimental data for 
$p p \rightarrow d \pi^+ $ are corrected for 
these electromagnetic effects, then both reactions 
near threshold can be viewed
as a determination of the single constant $\alpha$.
The threshold behavior of the reaction $pp\rightarrow pn\pi^+$
was estimated by Gell-Mann and Watson \cite{GW54}, who predicted an energy
dependence $\sigma\propto\eta^{4}$ under the approximation
of zero-range nuclear forces.

\begin{figure}
\epsfxsize=13.0cm
\epsfysize=4.0cm
\centerline{\epsffile{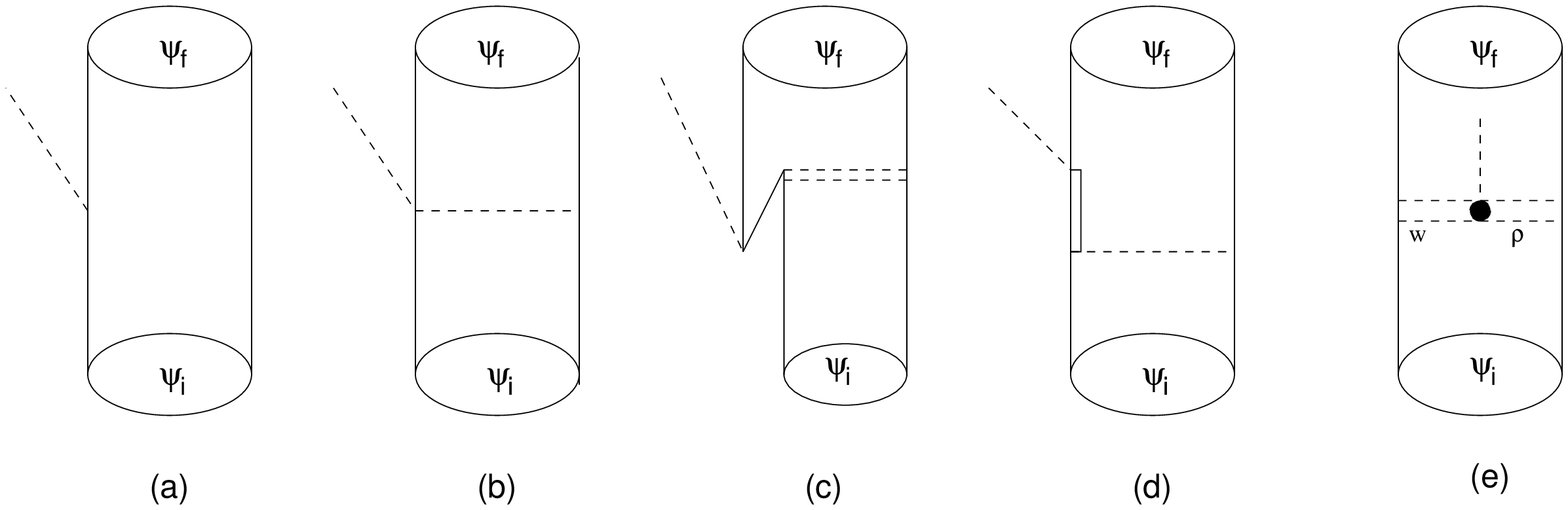}}
\vspace{.5cm}
\caption{ Various contributions to the $p p \rightarrow 
d \pi^+, pn\pi^+$ reactions. A  single (double) 
solid line stands for a nucleon (delta) and a single 
(double) dashed line represents a pion (sigma, omega, 
rho); $\Psi_i$ ($\Psi_f$) is the wave function for the 
initial (final) state.
\label{Fig.1}}
\end{figure}

Early theoretical analyses of the reaction  
$p p \rightarrow d \pi^+$, like in ours, 
split it into a kernel ---where
the two-nucleon system emits the pion--- and  
effects of initial- and final-state 
interactions ---taken into account by using
deuteron and scattering wave 
functions that satisfy the Schr\"odinger 
equation with a given $NN$ potential. 
The numerical results of 
the early analyses \cite{KR66} were that 
this reaction is dominated by the Weinberg-Tomozawa (WT) term
(Fig.~\ref{Fig.1}b). 
The main competitor process 
was thought to be the impulse term (IA)
(Fig.~\ref{Fig.1}a). The early 
estimate of this IA gave a small 
contribution \cite{KR66} due to a 
cancellation in the matrix elements between 
the $S$ and $D$ waves of the deuteron final 
state. It was believed that the WT 
term alone could account for 
the essential features of the data.

With the advent of accelerators capable of 
producing intense high-quality beams of 
protons with GeV energies and
very precise detecting systems, good and 
accurate data for $np\rightarrow d \pi^0$ \cite{triumf}, 
$p p \rightarrow d \pi^+$ \cite{cosy,iucf},
and $p p \rightarrow pn \pi^+$ \cite{steve1,steve2}
near threshold became available. 
For the deuteron channels, the new data confirm the threshold behavior
(\ref{snp}) \cite{triumf}, and
show that 
the theoretical calculations started by Koltun and Reitan were 
in the right direction. 
More recent calculations of the WT and IA mechanisms 
\cite{niskanen,Han+98},
however, indicated that their strength might not be sufficient
to explain the new data.
As for the unbound final state, the 
recent measurements in the Indiana Cooler \cite{steve2} give
\begin{equation}
|a_1|^2\propto \eta^{3.2}.
\label{apot}
\end{equation}
A calculation \cite{unpharry}
using IA, on-shell pion rescattering,
and $\sigma,\omega$ Z-graphs finds good agreement with data in all
$pp\pi^0$, $pn\pi^+$, and $d\pi^+$ channels with a soft $\pi N$
form factor.

Our motivation here is to examine what a chiral power counting 
suggests for the pion production in the $pp\rightarrow d\pi^+$
and $pp\rightarrow pn\pi^+$ reactions near threshold. 
For the $d \pi^+$ reaction, we will compute $\sigma_{pp}$ 
but will disregard the Coulomb interaction in both the initial 
state ---which is not so important due to the not-so-small initial 
energy--- and final state. This means we do not calculate $C_0(\eta)$,
and can only compare with Coulomb-corrected $pp$ data.
We will see that $\sigma_{pp}$ is indeed approximately linear with $\eta$
at threshold, so our result can be considered a calculation of $\alpha$ in 
Eq.~(\ref{spp}). Alternatively, it is a calculation of $2 \sigma_{np}$.
For the $pn\pi^+$ final state, we will also disregard the Coulomb 
interaction and see that our cross section has an energy dependence
similar to the data.
Moreover, we 
will show that the WT term is indeed the dominant one, not only for the
$^3S_1$ $pn\pi^+$ final state but also for the deuteron channel, 
which is consistent with the early analyses for 
this final state.
The IA term is smaller than WT because of cancellations 
between different regions in coordinate space, and
between the $S$ and $D$ waves of the deuteron final 
state. The same cancellations affect the contribution from
an explicit $\Delta$ in the intermediate state
(Fig.~\ref{Fig.1}d), more so in the $d \pi^+$  channel than in the 
$pn \pi^+ $ channel. The relative smallness of this term here 
reduces the effects of using different $NN$ potentials
compared to the $pp\rightarrow pp\pi^0$ reaction.
However the delta contribution is affected by a number of
uncertainties. We include also the same sub-leading terms that proved 
important in the $pp\rightarrow pp\pi^0$ reaction: isospin-independent 
rescattering (ST, Fig.~\ref{Fig.1}b), 
and heavy meson exchange simulating short-range
mechanisms ($\sigma,\omega$ and $\rho-\omega$, Fig.~\ref{Fig.1}c,e). 
The sub-leading Galilean correction to the WT term (GC)
is included as well. These contributions are all relatively small.
We will see that, if we neglect the uncertain
$\Delta$ contributions,
each of the the cross-sections is  underestimated by a factor of $\sim 2$.

In Section II we discuss the power counting and the chiral Lagrangian.
In Section III the kernel is obtained.
In Section IV the calculation of the cross-sections is outlined;
some technical details are relegated to an Appendix.
Section V describes our input and discusses our results.
An outlook is presented in Section VI.

\section{Implementing $\chi$PT}

Near threshold for pion production the total energy of the two colliding
nucleons is of order  
$2 m_N + m_\pi$,  
so that the center-of-mass initial kinetic energy of each nucleon is
$m_\pi/2$.  This energy is smaller than the nucleon mass,
so we can use a non-relativistic framework.
The nonrelativistic kinetic 
energy formula holds; the mass $2m_N$ plays no dynamical
role,    
and the typical momentum of real and virtual
particles involved in this process is 
$p_{\rm typ} \sim \sqrt{m_N m_\pi}$.
This requires
some adaptation of the usual effective theory ideas
which have been developed for momenta typical
of most nuclear systems, $Q \sim m_{\pi}$.
Because the mass difference between the delta isobar and the nucleon,
$\delta= m_\Delta - m_N$, is numerically of order of the typical
excitation energy we are interested in, $m_\pi$,
the $\Delta$ must be included explicitly  as a
degree of freedom in the Lagrangian. 
On the other hand, $\sqrt{m_N m_\pi}$
is smaller than the characteristic mass scale of QCD,
$M_{QCD}\sim 1$GeV, at least in the chiral limit $m_\pi\rightarrow 0$,
so that the contribution of other states (the Roper, the $\rho$ meson, 
{\it etc.}) can be buried in short-range interactions.

We thus seek a theory of non-relativistic nucleons and deltas 
interacting with pions  that is consistent with the symmetries of QCD.
$\chi$PT is implemented via the most general 
Lagrangian involving these degrees of freedom aided by power-counting 
arguments.  The seminal idea was contained 
in a paper by Weinberg \cite{W79}.  This 
idea was developed systematically for 
interactions of mesons \cite{GL84} and 
for interactions of mesons with a baryon 
\cite{JM91,BKM95}. The generalization of 
these techniques to describe properties 
of more than one  baryon was also due to 
Weinberg \cite{W90} and was carried out 
in detail in Ref. \cite{OK92,ORV94,V94}; 
see also Refs. \cite{vKolckn,KSW,monster}. 

When dealing with typical momenta of order
$Q \sim m_{\pi}$, the usual power counting
suggests that we order terms in the chiral
Lagrangian according to the ``index''
$\Delta=d+\frac{f}{2}-2$,
where $d$ is the sum of the number of derivatives,
the number of powers of $m_{\pi}$, 
and the number of powers of $\delta$;
and $f$ is the number of fermion field operators.
In the following we will consider interactions
with $\Delta$ up to 4.

The Lagrangian with $\Delta= 0$ 
for each interaction \cite{W79,GL84,JM91,BKM95,W90,ORV94,V94} is
\newcommand{\boldpi}{\mbox{\boldmath $\pi$}}
\newcommand{\boldtau}{\mbox{\boldmath $\tau$}}
\newcommand{\boldT}{\mbox{\boldmath $T$}}
\begin{eqnarray}
 {\cal L}^{(0)} & = & 
          \frac{1}{2}(\dot{\boldpi}^{2}-(\vec{\nabla}\boldpi)^{2})
          -\frac{1}{2}m_{\pi}^{2}\boldpi^{2} \nonumber   \\
    &   & +N^{\dagger}[i\partial_{0}-\frac{1}{4 f_{\pi}^{2}} \boldtau \cdot
         (\boldpi\times\dot{\boldpi})]N +\frac{g_{A}}{2 f_{\pi}} 
         N^{\dagger}(\boldtau\cdot\vec{\sigma}\cdot\vec{\nabla}\boldpi)N
                                               \nonumber \\
    &   & +\Delta^{\dagger}[i\partial_{0}- \delta]\Delta 
          +\frac{h_{A}}{2 f_{\pi}}[N^{\dagger}(\boldT\cdot
          \vec{S}\cdot\vec{\nabla}\boldpi)\Delta +h.c.] +\cdots \, , 
\label{la0}
\end{eqnarray} 
\noindent
where $f_{\pi}=93$ MeV is the pion decay constant,
$\delta = m_{\Delta}-m_{N}$ is the isobar-nucleon mass difference,
$g_{A}$ is the axial-vector coupling of the nucleon, $h_{A}$
is the~$\Delta N \pi$ coupling, and $\vec{S}$ and $\boldT$ are the 
transition spin and isospin matrices, normalized such that
\begin{eqnarray}
  S_{i}S^{+}_{j} & = & \frac{1}{3} (2\delta_{ij} - 
              i\varepsilon_{ijk} \sigma_{k})    \label{S1}   \\
  T_{a}T^{+}_{b} & = & \frac{1}{3} (2\delta_{ab} 
              - i\varepsilon_{abc} \tau_{c}).   \label{S2}
\end{eqnarray}
\noindent
Notice that we defined the fields $N$ and $\Delta$ in
such a way that there is no factor of $\exp (-i m_N t)$ in their
time evolution. Hence $m_N$ does not appear explicitly at this index,
corresponding to static baryons.  
We also wrote ${\cal L}^{(0)}$ in the rest frame
of the baryons, which is the natural choice. (Galilean invariance
will be ensured by including terms with additional derivatives.) 
Chiral symmetry determines the coefficient of the so-called
Weinberg-Tomozawa term 
($N^{\dagger}\boldtau \cdot(\boldpi\times\dot{\boldpi})N$)
but not of the single-pion interactions ($g_A, h_A$).
              
The Lagrangian with $\Delta=1$ is 
\cite{JM91,BKM95,ORV94,V94},
\begin{eqnarray}
 {\cal L}&^{(1)}& 
        =\frac{1}{2m_{N}}[N^{\dagger}\vec{\nabla}^{2}N
        +\frac{1}{4 f_{\pi}^{2}}(iN^{\dagger}\boldtau\cdot
        (\boldpi\times\vec{\nabla}\boldpi)\cdot\vec{\nabla}N + h.c.)] 
                                       \nonumber \\
  &   & +\frac{1}{f_{\pi}^{2}}N^{\dagger}[(c_2 +c_3 - \frac{g_A ^2}{8 m_{N}})
        \dot{\boldpi}^{2} -c_3 (\vec{\nabla}\boldpi)^{2} 
        -2c_1 m_{\pi}^{2} \boldpi^{2} -
        \frac{1}{2} (c_4 + \frac{1}{4m_{N}}) 
        \varepsilon_{ijk} \varepsilon_{abc} \sigma_{k} \tau_{c} 
        \partial_{i}\pi_{a}\partial_{j}\pi_{b}]N \nonumber   \\  
  &   & +\frac{\delta m_{N}}{2} N^{\dagger}[\tau_{3}-\frac{1}{2 f_{\pi}^{2}}
        \pi_3 \boldpi\cdot\boldtau]N  
        +\frac{1}{2m_{N}}\Delta^{\dagger}[\vec{\nabla}^{2} +\cdots]\Delta 
                                                               \nonumber \\
  &   & -\frac{g_{A}}{4 m_{N} f_{\pi}}[iN^{\dagger}\boldtau\cdot\dot{\boldpi}
        \vec{\sigma}\cdot\vec{\nabla}N + h.c.]             
        -\frac{h_{A}}{
        2 m_{N} f_{\pi}}[
        iN^{\dagger}\boldT\cdot\dot{\boldpi}\vec{S}\cdot\vec{\nabla}
        \Delta + h.c.]                       \nonumber \\
  &   & -\frac{d_1}{f_{\pi}} 
        N^{\dagger}(\boldtau\cdot\vec{\sigma}\cdot\vec{\nabla}\boldpi)N\,
        N^{\dagger}N
        -\frac{d_2}{2 f_{\pi}} \varepsilon_{ijk} \varepsilon_{abc} 
        \partial_{i}\pi_{a}  
        N^{\dagger}\sigma_{j}\tau_{b}N\, N^{\dagger}\sigma_{k}\tau_{c}N 
        +\cdots \, ,             \label{la1}
\end{eqnarray}
\noindent
where the $c_{i}$'s are coefficients of ${\cal O}(1/M)$, 
$\delta m_{N} \sim m_d -m_u$ is the quark mass difference contribution to 
the neutron-proton mass difference,
and the $d_{i}$'s are coefficients of ${\cal O}(1/f_\pi^2 M)$. 
These seven numbers are not fixed by chiral symmetry, but it is important 
to point out that Galilean invariance requires that the other coefficients 
explicitly shown above be related to those appearing in ${\cal L}^{(0)}$. 
This in particular
fixes the strength of 
the single-pion interactions in terms of the 
lowest-order coefficients $g_A$ and $h_A$, and of the common mass $m_{N}$. 

The Lagrangian with $\Delta=2$ is
\begin{eqnarray}
{\cal L}^{(2)} & = & \frac{d_1^{\prime}+e_1}{2 m_N f_{\pi}} 
                 [iN^{\dagger}\boldtau\cdot\dot{\boldpi}
                 \vec{\sigma}\cdot\vec{\nabla}N\, N^{\dagger}N + h.c.] 
                                                   \nonumber \\
           &  & -\frac{e_1}{2 m_N f_{\pi}}
                 [iN^{\dagger}\boldtau\cdot\dot{\boldpi}\vec{\sigma}N\,
                 \cdot N^{\dagger}\vec{\nabla}N + h.c.] \nonumber \\
           &  & +\frac{e_2}{2 m_N f_{\pi}}
                 [N^{\dagger}\boldtau\cdot\dot{\boldpi}\vec{\sigma}\times
                 \vec{\nabla}N\,
                 \cdot N^{\dagger}\vec{\sigma}N + h.c.] 
                 +\cdots \, ,                  \label{la2}
\end{eqnarray}
\noindent
where the $e_i$'s are other coefficients of ${\cal O}(1/f_\pi^2 M)$. 

Among the Lagrangians with higher indices, we find 
\begin{equation}
{\cal L}^{(4)} =
\frac{g}{2 m_N f_{\pi}}
                 [iN^{\dagger}\boldtau\cdot\dot{\boldpi}
\vec{\sigma}\cdot\vec{\nabla}\vec{\nabla}N \cdot N^{\dagger}\vec{\nabla}N 
+ h.c.] 
              +\cdots \, ,  \label{la3}
\end{equation}
\noindent
where $g$ is
a coefficient of 
${\cal O}(1/f_\pi^2 M^3)$.

The two nucleons in the $N N \rightarrow NN \pi$ reaction 
can interact repeatedly by the exchange of mesons of momenta 
$Q\sim m_\pi$ before and after the emission of the pion.
We account for this through the iteration of a potential,
which produces initial and final wavefunctions that
differ from the free ones.
The emission of the pion, on the other hand,
involves the larger momentum $p_{\rm typ} \sim \sqrt{m_N m_\pi}$.
The sub-diagrams that involve such typical momentum
form the kernel of ``irreducible diagrams'',
which is evaluated between wavefunctions.

The unusually high momentum in the kernel 
requires modification in the usual power counting.
While in the usual power counting energy and momenta are counted
as equal, here energies are $\sim m_\pi$ but momenta
$\sim \sqrt{m_N m_\pi}$. This changes the usual correspondence
between index and order.

The leading nucleon propagator, for example, includes both the
static term in (\ref{la0}) and the kinetic term in  (\ref{la1})
at the same order in power counting for pion production, 
contrary to the situation in 
$\chi$PT applied to one-baryon systems. The nucleon
propagator then is $\sim 1/m_\pi$. 
Relativistic corrections $p^4/8m_N^3 + \ldots$ are relatively smaller by 
$\sim m_\pi/m_N$ and can be considered higher-order insertions.
Note that this is completely consistent with 
our decomposition of the full amplitude into a kernel
and wavefunctions obtained from a Schr\"odinger equation,
and,
contrary to what is stated in Ref. \cite{cagadadoulf},
does {\it not} imply the need of a relativistic framework.
The delta propagator differs from the nucleon 
by the presence of the mass difference $\delta\sim 2m_\pi$.
As pointed out first in Ref. \cite{bira},
the delta propagator is then actually $\sim -1/m_\pi$ and tends to 
interfere destructively with the nucleon contributions to the kernel.
The pion propagator, on the other hand,
is $\sim 1/m_\pi m_N$.
In interactions, each time derivative is associated with a factor
of $m_\pi$, while a space derivative a factor of $\sqrt{m_N m_\pi}$.
Finally, a loop brings a $(m_N m_\pi)^{3/2}m_\pi/(4\pi)^2$.

A detailed analyses of various contributions 
can found in Ref. \cite{bira}. 
The new elements here are those associated with 
the 
isospin-dependent WT  
pion rescattering. 
The order of the WT 
contribution is evaluated as follows. The
term proportional to $\partial_0\bbox{\pi}$
of Eq.~(\ref{la0}) yields an 
explicit factor of $m_\pi/f_\pi^2$; the
pion-nucleon interaction provides a factor 
of $\sqrt{m_\pi\,m_N}/f_\pi$, and the 
pion propagator $(m_\pi\,m_N)^{-1}$. The 
total net result is 
of order ${\cal O}(f_{\pi}^{-3} \sqrt{m_\pi /M})$,
which is the same order of the impulse approximation 
and the delta contribution \cite{bira}.
Therefore due to power counting arguments, 
we expect that contribution of WT, IA, and $\Delta$ terms to have the same 
importance; they constitute our leading order.
We will also include 
other terms which are of order 
${m_\pi /M}$ or higher relative to the leading one: 
the isospin-independent pion rescattering  
(SG), 
the Galilean correction to the WT
term (GC), and contact terms, modeled by 
the heavy meson exchange ($\sigma$, $\omega$ and $\rho-\omega$).

\section{The Kernel}

We now obtain the explicit forms of the various 
contributions by evaluating the most important 
irreducible diagrams in momentum space.
Our notation is as follows:
$\omega_q^2 = \vec{q}^{\: 2} + m_{\pi}^2$ 
is the energy of the (on-shell) pion produced 
with momentum $\vec{q}$ in the center of mass;
$\vec{p}$ ($\vec{p'}$) 
is the center-of-mass momentum of the 
incoming (outgoing) nucleon labeled 
``1'' (those of nucleon ``2'' are opposite);
$\vec{k}= \vec{p} -\vec{p'}$ 
($k^0= m_\pi/2$)
%
is the momentum (energy) transferred;
$\omega_k^2 = \vec{k}^2 + m_{\pi}^2$;
$\vec{P}= \vec{p} +\vec{p'}$; 
$\vec{\sigma}^{(i)}$ 
is the spin of proton $i$;
$\vec{\Sigma}_a= \vec{\sigma}^{(1)}\tau_a^{(1)} - 
\vec{\sigma}^{(2)}\tau_a^{(2)}$ 
where $a$ is the isospin of emitted pion;
and $T(\vec{k})\equiv \vec{\sigma}^{(1)}\cdot\vec{k} \;
\vec{\sigma}^{(2)}\cdot\vec{k}$.  
We define the $T$-matrix in terms of the 
$S$-matrix via $S=1+iT$. 

According to the previous discussion, we 
expect the leading contributions
to arise {}from the diagrams in 
Figs.~\ref{Fig.1}a,b, and d.
In the case of the Weinberg-Tomozawa 
diagram (Fig.~\ref{Fig.1}b), we get

\begin{equation}
T^{WT}_a=-\frac{g_A}{4f_\pi^3}\,\epsilon_{abc}\tau_b^{(1)}\tau_c^{(2)}\,
\frac{\omega_q+k^0}{w_k^2-(k^0)^2}\;\vec{S}\cdot\vec{k}\; .
\label{WT}
\end{equation}

\noindent
The Galilean correction to WT contribution is smaller by a factor of 
$m_\pi/M$ and is given by 

\begin{equation}
T^{GC}_a=-\frac{g_A}{8m_Nf_\pi^3}\,\epsilon_{abc}\tau_b^{(1)}\tau_c^{(2)}\,
\frac{1}{w_k^2-(k^0)^2}\;\vec{k}\cdot\vec{P}\;\vec{S}\cdot\vec{k}\; .
\label{GC}
\end{equation}

The impulse term (Fig.~\ref{Fig.1}a)
is discussed in detail in \cite{bira}.
The principle of irreducibility says 
that we have to redraw this
diagram. Since the outgoing pion 
carries an energy of the order of 
the pion mass, the energies of the 
$NN$ intermediate state before and 
after pion emission differ by 
$\sim m_\pi$. Therefore both of the 
intermediate states cannot simultaneously be 
within $\sim m_\pi^2/m_N$ of being on-shell:
at least one intermediate state, before 
or after emission, is off shell by $\sim m_\pi$. 
This single, relatively-high-momentum 
($\sim \sqrt{m_{\pi} m_N}$) pion exchange 
must therefore be included in the 
irreducible class of operators for 
our process (unlike the usual case).
All other initial- and final-state 
interactions will be considered 
reducible and included in the wave 
functions. 
Therefore, in the case of pion exchange with a 
nucleon in the intermediate state 
we get
\begin{equation}
T^{IA}_a=\frac{ig_A^3}{8 m_{N} f_{\pi}^3}\; \frac{1}{\omega_k^2-(k_0)^2}
       \left[\vec{\Sigma}_a\cdot\vec{p'}\; T(\vec{k})- 
        T(\vec{k})\, \vec{\Sigma}_a\cdot\vec{p}\;\right]\, ,
\label{IA}
\end{equation}
\noindent which is listed for comparative 
purposes only. We will actually calculate 
the impulse approximation directly 
from Eq.(6) in Ref.~\cite{bira}, in the same
fashion as done by 
Koltun and Reitan \cite{KR66}.

Recoil corrections are expected to be smaller
by a factor of $m_{\pi}/M$. Since the impulse contribution will prove
to be small, we can ignore the recoil contributions in this first approach.

The $\Delta$ contribution (Fig.~\ref{Fig.1}d)
to the kernel is given by 
\begin{eqnarray}
T^{\Delta}_a & = & 
       \frac{-ig_A h_A^2}{18 m_{N} f_{\pi}^3}\; \frac{1}{\omega_k^2-(k^0)^2}
       \frac{\omega_q}{\delta^2-\omega_q^2} \times  \nonumber \\
  & &   \left[\left(\vec{k}^2 \omega_q - \vec{k}\cdot\vec{P}\;\delta\right)   
        \vec{\Sigma}_a\cdot \vec{k}
       +\frac{i}{2} \omega_q \left(\tau_a^{(1)}\;\vec{\sigma}^{(1)}
        \cdot\vec{k}\;\vec{\sigma}^{(2)}\cdot(\vec{P}\times\vec{k})           
        -\tau_a^{(2)}\;\vec{\sigma}^{(1)}\cdot(\vec{P}\times\vec{k})         
        \;\vec{\sigma}^{(2)}\cdot\vec{k}\right)\right]\nonumber \\
  & &   +i\epsilon_{abc}\tau_b^{(1)}\tau_c^{(2)}\left[\left(\omega_q\,\vec{k}
        \cdot\vec{P}-\delta\,\vec{k}^2\right)\vec{S}\cdot\vec{k}-
        \frac{i}{4}\delta\left(\vec{\sigma}^{(1)}
        \cdot\vec{k}\;\vec{\sigma}^{(2)}\cdot(\vec{P}\times\vec{k})           
        +\vec{\sigma}^{(1)}\cdot(\vec{P}\times\vec{k})         
        \;\vec{\sigma}^{(2)}\cdot\vec{k}\right)\right].
\label{Delta}
\end{eqnarray}
\noindent
Results similar to Eqs. (\ref{IA}) 
and (\ref{Delta}) follow for shorter-range 
terms where the two nucleons exchange a heavier meson rather
than a pion.
In the case of a nucleon intermediate state,
such a contribution is automatically included in the potential.
In any reasonable model, the contributions from a delta
intermediate state
turn out to be smaller than those 
in diagram of Fig.~\ref{Fig.1}d. For example,
they could arise {}from $a_1$ exchange, but 
then the relatively high $a_1$ mass suppresses 
this contribution;
the contribution from the $\rho$ which is formally
of higher-order is likely to be more important.
Since, as we are going to see, the delta contribution 
from pion exchange is not large,
we will not go into such detailed analysis for the purpose of estimating 
the effect of the $\Delta$: we use 
Eq.(\ref{Delta}). We will further discuss the uncertainties related to
the delta contribution below.

There are other corrections of order $m_{\pi}/M$ 
compared to the leading terms.
Fig.~\ref{Fig.1}b represents also isospin-independent rescattering:
\begin{eqnarray}
T^{ST}_a&=&i\frac{g_A}{f_{\pi}^3}\, \frac{1}{\omega_k^2 -(k^0)^2}
       \left\{\left[\left(c_2 +c_3- \frac{g_A^2}{8 m_N}\right) k^0 
       \omega_q -2c_1 m_{\pi}^2\right]\vec{\Sigma}_a\cdot\vec{k}
       \right..\nonumber \\
 & -& \left.\frac{\delta m_N}{8}\left[\delta_{3a}\vec{\tau}^{(1)}\cdot
      \vec{\tau}^{(2)}\,\left(\vec{\sigma}^{(1)}-\vec{\sigma}^{(2)}\right)
      +\vec{\sigma}^{(1)}\tau_3^{(1)}\tau_a^{(2)}-
       \vec{\sigma}^{(2)}\tau_a^{(1)}\tau_3^{(2)}\right]\cdot\vec{k}\right\}.
\label{ST}
\end{eqnarray}

The short-range mechanisms provided by the $\Delta=2, 3, 4$ Lagrangians
involve several unknown constants. 
Chiral symmetry tells us nothing about the strength of these coefficients.
We can use data to determine some of them.
Alternatively,
we can use a model to determine these coefficients and then try to explain
the experimental results. Here we use the mechanism first proposed
by Lee and Riska \cite{LR92} and by Horowitz {\it et al} \cite{HGM94},
where the short-range interaction is supposed to originate {}from
Z-graphs with $\sigma$ and $\omega$ exchanges, as shown in Fig.~\ref{Fig.1}c. 
In this case,
\begin{eqnarray}
T^{\sigma, \omega}_a & = & - \frac{i\, g_A}{4 f_{\pi} m_N^2}\; \omega_q 
      \left[\left(\frac{g_\sigma^2}{\vec{k}^2 + m_{\sigma}^2}+ 
                   \frac{g_\omega^2}{\vec{k}^2 + m_{\omega}^2}\right)
              \vec{\Sigma}_a\cdot\vec{P}\right. \nonumber\\ 
  &   &\left. - i\,\frac{g_\omega^2 (1+C_\omega)}{\vec{k}^2 + m_{\omega}^2} 
         \;\vec{\sigma}^{(1)}\times\vec{\sigma}^{(2)}\cdot\vec{k}
    \, \left(\tau_a^{(1)}+\tau_a^{(2)}\right)\right] ,
\label{sigom}
\end{eqnarray}
where $m_\sigma$ ($m_\omega$) and $g_\sigma$ ($g_\omega$) are 
the mass and the vector coupling to nucleons of the 
$\sigma$ ($\omega$) meson, 
and $C_\omega$ denotes the ratio of tensor to vector coupling for the 
$\omega$ meson. 
In the case of $pp\rightarrow pp \pi^0$,
the contribution due to $\rho-\omega$ exchange (Fig.~\ref{Fig.1}e)
is not negligible
\cite{KMR96}, so we also include it here in order to get an estimate
of the convergence of our expansion. This contribution leads to
\begin{eqnarray}
T^{\rho, \omega}_a & = & - \frac{i\, g_\rho g_\omega g_{\pi\rho\omega}}
{4 m_N^2}\; \frac{\omega_q}{m_\omega} 
      \left(\frac{g_\rho^2}{\vec{k}^2 + m_{\rho}^2}\cdot 
                   \frac{g_\omega^2}{\vec{k}^2 + m_{\omega}^2}\right) \nonumber \\
  & & \left\{\left(2+C_\rho+C_\omega\right)\left(\vec{k}^2
              \vec{\Sigma}_a\cdot\vec{P}-\vec{k}\cdot\vec{P}\vec{\Sigma}_a
\cdot\vec{k}\right)\right. - \nonumber\\
  &   &\left. -i\,\left(1+C_\rho\right)\left(1+C_\omega\right)
        \vec{k}^2\,\vec{\sigma}^{(1)}\times\vec{\sigma}^{(2)}\cdot\vec{k}
    \, \left(\tau_a^{(1)}+\tau_a^{(2)}\right)\right\},
\label{rhoom}
\end{eqnarray}
\noindent 
where $g_{\pi\rho\omega}$ is the $\pi \rho \omega$ coupling \cite{KMR96} and
the other coefficients have the same meaning as in the previous equation.
At momenta much smaller than the meson masses,
these short-range contributions are indeed 
contact interactions.

\section{Cross-section}

\subsection{The $pp\rightarrow d\pi^+$ reaction}

We are concerned with evaluating the matrix elements of the above 
operators between the initial $^3P_1$ and the deuteron final 
wave functions.
To evaluate the influence of the potential in the amplitudes, 
we use Reid93 \cite{Reid93}
and Argonne V18 \cite{v18} potentials which, for a given $pp$ channel, 
are local potentials. Thus we evaluate the operators between coordinate 
space initial ($i$) and final ($f$) 
wave functions expressed by

\begin{equation}
\langle \vec{r}|i \rangle =\mp{\sqrt{2}\over pr} i \, u_{1,1}(r)e^{i
\delta_{1,1}} \left(\sqrt{2\pi}\;\sqrt{3}\right)\, |^3P_1\rangle \;,
\end{equation}
\noindent 
where $-$ ($+$) is for third-component of the angular momentum 
$M_J=+1$ ($-1$),  
and
\begin{equation}
\langle \vec{r}|f \rangle ={1\over r} \left[u(r)\;|^3S_1\rangle 
                           +w(r)\;|^3D_1\rangle \right]\;,
\end{equation}
\noindent where the deuteron wave functions are normalized as
\begin{equation}
\int_0^\infty dr\left[u^2(r)+w^2(r)\right]=1\;.
\end{equation}
Here and in the following
the spin angular part is expressed, as usual, as

\begin{equation}
| ^{2S+1}L_J \rangle = \sum_{m_L\,m_S}\langle L\,m_L\,S\,m_S|J\,m_J\rangle
                       Y_L^{m_L}(\hat{\bf r})\chi_S^{m_S}\;,
\end{equation}

\noindent 
where $Y_L$ is the spherical harmonic function.

We convert the operators of Eqs.~(\ref{WT}-\ref{rhoom}) to
configuration space by inverting the Fourier transforms.
The resulting operators can then be used in configuration-space matrix
elements. 
We define the matrix 
elements of the operators of Eqs.~(\ref{WT}-\ref{rhoom}) as  
\begin{eqnarray}
{\cal M}^{X} & = & \langle f|T^{X}|i\rangle  \nonumber \\
      & = & \mp{\sqrt{12\pi}\;i\over p}
            e^{i\delta_{1,1}} 
         \int_0^\infty dr\left[\frac{u(r)}{\sqrt{m_\pi}}\,H^{X}_S(r)+
                    \frac{w(r)}{\sqrt{2m_\pi}}\,H^{X}_D(r)\right]\,u_{1,1}(r),
\label{Mdef}
\end{eqnarray}

\noindent
where $X$ represents $WT$, $IA$, {\it etc.}
$H^{X}_{S,D}(r)$ is the corresponding operator, 
obtained using the matrix elements given in the Appendix. 
To compare results before the cross-section evaluation, we define
dimensionless amplitudes $J^X$, where $X=WT,IA,\Delta,$ {\it etc.}
via

\begin{equation}
{\cal M}^{X} =  \mp \,\sqrt{12\pi}\;i\;e^{i\delta_{1,1}}
         \frac{g_A}{f_\pi^3}\;\cdot\frac{m_\pi}{4\pi}\, \sqrt{\frac{2}{3}} 
         \; J^{X}.
\label{Mdef2}
\end{equation}
We will plot $J^{X}$ in terms of $\eta$ to evaluate the energy dependence 
of the amplitude, 
and $dJ^X/dr$ in terms of $r$ to study the $r$ dependence of the integrand.

The contribution from the Weinberg-Tomozawa term is given by 
\begin{equation}
H^{WT}_S(r)=H^{WT}_D(r)={g_A \over 4 f^3_\pi}\frac{3 m_\pi}{4\pi}
\sqrt{\frac{2}{3}}\;{(1+\tilde{m}_\pi r)\over r^2} e^{-\tilde{m}_\pi r},
\label{wt}
\end{equation}

\noindent where $\tilde{m}_\pi=\sqrt{\frac{3}{4}}m_\pi$. 
The Galilean correction to the WT term is given by
\begin{eqnarray}
H^{GC}_S(r)&=&{g_A \over 4 f^3_\pi}\frac{3 m_\pi}{4\pi}\sqrt{\frac{2}{3}}\; 
\frac{m_\pi}{4m_N}\left[A_{GC}(r)+B_{GC}(r)+C_{GC}(r)\right], 
\label{gcs}\\
H^{GC}_D(r)&=&{g_A \over 4 f^3_\pi}\frac{3 m_\pi}{4\pi} \sqrt{\frac{2}{3}}\;
\frac{m_\pi}{4m_N}\left[A_{GC}(r)+B_{GC}(r)+D_{GC}(r) \right],
\label{gcd}
\end{eqnarray}
\noindent where 
\begin{eqnarray}
A_{GC}(r)&=&{(1+\tilde{m}_\pi r)\over r^2} e^{-\tilde{m}_\pi r}, \\
B_{GC}(r)&=&-2 \frac{ e^{-\tilde{m}_\pi r}}{r}\left(1+\frac{2}{\tilde{m}_\pi r}
+\frac{2}{(\tilde{m}_\pi r)^2}\right)\frac{\partial}{\partial r}, \\
C_{GC}(r)&=&2 \frac{ e^{-\tilde{m}_\pi r}}{r^2}\left(1+\frac{4}{\tilde{m}_\pi r}
+\frac{4}{(\tilde{m}_\pi r)^2}\right), \\
D_{GC}(r)&=&2 \frac{ e^{-\tilde{m}_\pi r}}{r^2}\left(1+\frac{1}{\tilde{m}_\pi r}
+\frac{1}{(\tilde{m}_\pi r)^2}\right). 
\end{eqnarray}

The impulse approximation is given by
\begin{eqnarray}
H^{IA}_S(r)&=&{g_A \over f_\pi}\frac{m_\pi}{m_N}\sqrt{\frac{1}{3}}\; 
\left[\frac{1}{r}+\frac{\partial}{\partial r}\right], 
\label{ias}\\
H^{IA}_D(r)&=&{g_A \over f_\pi}\frac{m_\pi}{m_N}\sqrt{\frac{1}{3}}\; 
\left[-\frac{2}{r}+\frac{\partial}{\partial r}\right]. 
\label{iad}
\end{eqnarray}

The $\Delta$ resonance contribution 
be written as 
\begin{eqnarray}
H^{\Delta}_S(r)&=&\frac{1}{9m_N}\left({g_A \over f_\pi}\right)^3\left(
\frac{h_A}{g_A}\right)^2\frac{ m_\pi \tilde{m}_\pi^2}{(m_\pi-\delta)}
\sqrt{\frac{2}{3}}\;\frac{1}{4\pi}\left[A_{\Delta}(r)+B_{\Delta}(r)+
C_{\Delta}(r)\right], \label{dels}\\
H^{\Delta}_D(r)&=&\frac{1}{9m_N}\left({g_A \over f_\pi}\right)^3\left(
\frac{h_A}{g_A}\right)^2\frac{ m_\pi \tilde{m}_\pi^2}{(m_\pi-\delta)}
\sqrt{\frac{2}{3}}\;\frac{1}{4\pi}\left[A_{\Delta}(r)+B_{\Delta}(r)+
D_{\Delta}(r)\right],
\label{deld}
\end{eqnarray}
\noindent 
where 
\begin{eqnarray}
A_{\Delta}(r)&=&2{(1+\tilde{m}_\pi r)\over r^2} e^{-\tilde{m}_\pi r}, \\
B_{\Delta}(r)&=&-2\frac{ e^{-\tilde{m}_\pi r}}{r}
\left(1+\frac{1}{\tilde{m}_\pi r}+\frac{1}{(\tilde{m}_\pi r)^2}\right)
\frac{\partial}{\partial r}\;, \\
C_{\Delta}(r)&=&2 \frac{ e^{-\tilde{m}_\pi r}}{r^2}
\left(1+\frac{2}{\tilde{m}_\pi r}+\frac{2}{(\tilde{m}_\pi r)^2}\right), \\
D_{\Delta}(r)&=& \frac{ e^{-\tilde{m}_\pi r}}{r^2}
\left(-1+\frac{1}{\tilde{m}_\pi r}+\frac{1}{(\tilde{m}_\pi r)^2}\right). 
\end{eqnarray}

The isospin-independent seagull term is given by
\begin{equation}
H^{SG}_S(r)=H^{SG}_D(r)={g_A \over f^3_\pi}\frac{m_\pi^2}{4\pi}
\sqrt{\frac{2}{3}}F_{SG}\;{(1+\tilde{m}_\pi r)\over r^2} e^{-\tilde{m}_\pi r},
\label{sg}
\end{equation}
\noindent where
\begin{equation}
F_{SG}=4c_1+{\delta m_N\over 4 m_\pi^2}-\left(c_2+c_3-{g_A^2\over8m_N}\right).
\end{equation}

We also include the effects of the $\sigma$ and $\omega$ exchange and 
$\rho-\omega$ term. 
The result for the $\sigma$ and $\omega$ exchange is
\begin{equation}
H^{\sigma, \omega}_S(r)=\frac{g_A}{f_\pi} {m_\pi \over\,m_N^2}
              \sqrt{\frac{2}{3}}
              \left\{ \left[f_\sigma(r)+ f_\omega(r)\right]
              \left({\partial\over\partial r}+{1\over r}\right)
              + \frac{1}{2}\frac{\partial}{\partial r}\left[
               f_\sigma(r)+ f_\omega(r)\right]\right\}\, ,
\label{sig-omS}
\end{equation}
\noindent and
\begin{equation}
H^{\sigma, \omega}_D(r)=\frac{g_A}{f_\pi} {m_\pi \over\,m_N^2}
              \sqrt{\frac{2}{3}}
              \left\{ \left[f_\sigma(r)+ f_\omega(r)\right]
              \left({\partial\over\partial r}-{2\over r}\right)
              + \frac{1}{2}\frac{\partial}{\partial r}\left[
               f_\sigma(r)+ f_\omega(r)\right]\right\}\, ,
\label{sig-omD}
\end{equation}
\noindent
where the function $f_h(r)$ accounts for exchange of the meson $h$ between 
nucleons,
\begin{equation}
f_h(r)={g_h^2\over 4\pi}{e^{-m_h r}\over r}.
\end {equation}
\noindent
We follow Ref.\cite{HGM94} by including the effects of form factors 
in these heavy-meson contributions, as 
given in the Bonn potential \cite{machleidt}. 
For completeness, we repeat that monopole form factors are used 
at each meson vertex according to the replacement
\begin{equation}
g_h\to g_h{\Lambda_h^2-m_h^2\over\Lambda_h^2-k_\mu k^\mu},
\label{formf}
\end{equation}
where $k_\mu$ is the transferred momentum and $\Lambda_h$ is the 
cutoff mass.

With $M$ denoting the common $\rho-\omega$ mass (780 MeV), 
$\rho-\omega$ exchange yields
\begin{eqnarray}
H^{\rho-\omega}_S(r)&=&\frac{g_\rho\,g_\omega\,g_{\pi\rho\omega}}{4m_N^2} 
              {m_\pi \over\,4\pi} \sqrt{\frac{2}{3}}
              \frac{2(2+C_\rho)}{M(\Lambda^2-M^2)}\times\nonumber \\
       & & \left\{[A_{\rho-\omega}(r; M, \Lambda)
                   -A_{\rho-\omega}(r; \Lambda, M)]
                     \left({\partial\over\partial r}+{1\over r}\right)\right.
            \nonumber \\
       & & \left. \;\; 
            -[B_{\rho-\omega}(r; M, \Lambda)
                   -B_{\rho-\omega}(r; \Lambda, M)]
                     \left({\partial\over\partial r}-{1\over r}\right)\right\},
\label{roS}
\end{eqnarray}
where
\begin{eqnarray}
A_{\rho-\omega}(r; m_1, m_2) &=&
   m_1 \, \frac{e^{-m_1r}}{m_1r}\left[ 4m_1^2\left(\frac{1}{m_1r}+
              \frac{1}{(m_1r)^2}\right)-(m_2^2-m_1^2)m_1r
              +3m_1^2+m_2^2\right],\nonumber \\
B_{\rho-\omega}(r; m_1, m_2) &=&
   m_1 \, \frac{e^{-m_1r}}{m_1r}\left[ 4m_1^2\left(1+\frac{3}{m_1r}+
              \frac{3}{(m_1r)^2}\right)-(m_2^2-m_1^2)(1+m_1r)\right].
\end{eqnarray}
$H^{\rho-\omega}_D(r)$ 
can be obtained making the replacement 
\[
\left(\frac{\partial}{\partial r}\pm\frac{1}{r}\right)\longrightarrow 
\left(\frac{\partial}{\partial r}\pm\frac{2}{r}\right)
\] 
above.

The final steps consist of computing the total matrix element ${\cal M}$,
\begin{equation}
{\cal M}={\cal M}^{WT}+{\cal M}^{GC}+{\cal M}^{IA}+{\cal M}^{ST}+{\cal M}^{\Delta}
+{\cal M}^{\sigma,\omega}+{\cal M}^{\rho-\omega},
\label{calm}
\end{equation}
squaring it, and integrating over the available phase space. 
We find
\begin{equation}
\sigma={1\over 16\pi}\;\frac{m_\pi}{p}\;E_d\,\omega_q\,\eta|{\cal M}|^2,
\label{cross}
\end{equation}
where $p$ is the magnitude of the center-of-mass 3-momentum,
\[
p=m_\pi\cdot \sqrt{\frac{m_N}{m_\pi}\cdot\sqrt{1+\eta^2}+
\frac{\eta^2\;m_N}{2M_d}+\frac{(M_d-2m_N)m_N}{m_\pi^2}},
\]
and $E_d$ is energy of the produced deuteron
of mass $M_d$, $E_d=\sqrt{M_d^2+q^2}$.


\subsection{The $pp\rightarrow pn\pi^+$ reaction}


As we did in the deuteron case, we evaluate the matrix elements
for the unbound final state using the same operators of 
Eqs.~(\ref{WT})-(\ref{rhoom}). We will consider here just the absolute
threshold limit where $L_{\pi(NN)}=0$. According to selection rules,
we have 2 channels, as stated in Tab.~\ref{Tab.01}: 
$^3P_1\rightarrow {^3S_1}$ ($T_i=1,\,T_f=0$) and 
$^3P_0\rightarrow {^1S_0}$ ($T_i=1,\,T_f=1$). Again, 
to evaluate the influence of the potential in the amplitudes, 
we use Reid93 \cite{Reid93}
and Argonne V18 \cite{v18} potentials which, for a given $pp$ or $pn$ 
channel, are local potentials. Thus we evaluate the operators between 
coordinate space initial ($i$) and final ($f$) 
wave functions for the two channels, expressed by

\begin{itemize}
\item $^3P_1\rightarrow$ $^3S_1$ channel:

\begin{eqnarray}
\langle \vec{r}|i \rangle &=& \mp{\sqrt{2}\over pr} i \, u_{1,1}(r)e^{i
\delta_{1,1}} \left(\sqrt{2\pi}\;\sqrt{3}\right)\, |^3P_1\rangle \;,
\\ \nonumber 
\langle \vec{r}|f \rangle &=& {1\over p'r} i \, u_{0,1}(r)e^{i
\delta_{0,1}} \sqrt{4\pi}\, |^3S_1\rangle \;,
\end{eqnarray}

\item $^3P_0\rightarrow$ $^1S_0$ channel:

\begin{eqnarray}
\langle \vec{r}|i \rangle &=& {\sqrt{2}\over pr} i \, u_{1,0}(r)e^{i
\delta_{1,0}} \sqrt{4\pi}\, |^3P_0\rangle \;,
\\ \nonumber 
\langle \vec{r}|f \rangle &=& {1\over p'r} i \, u_{0,0}(r)e^{i
\delta_{0,0}} \sqrt{4\pi}\, |^1S_0\rangle \;,
\end{eqnarray}

\end{itemize}

\noindent 
where $-$ ($+$) is for third-component of the angular momentum 
$M_J=+1$ ($-1$).

We convert the operators of Eqs.~(\ref{WT}-\ref{rhoom}) to
configuration space by inverting the Fourier transforms.
The resulting operators can then be used in configuration-space matrix
elements. 
We define the matrix 
elements of the operators of Eqs.~(\ref{WT}-\ref{rhoom}) as  

\begin{itemize}

\item $^3P_1\rightarrow$ $^3S_1$ channel

\begin{equation}
{\cal M}^{X} = \langle ^3S_1 | T^X | ^3P_1 \rangle =  
                 \mp \,4\pi\,\sqrt{3}\;i\;e^{i(\delta_{0,1}+\delta_{1,1})}
         \frac{g_A}{f_\pi^3}\;\cdot\frac{1}{4\pi}\, \sqrt{\frac{2}{3}} 
         \; J^{X}.
\label{Mdef3p1}
\end{equation}

\item $^3P_0\rightarrow$ $^1S_0$ channel

\begin{equation}
{\cal M}^{Y} = \langle ^1S_0 | T^Y | ^3P_0 \rangle =  
                 4\pi\,\sqrt{2}\;i\;e^{i(\delta_{1,0}+\delta_{0,0})}
         \frac{g_A}{f_\pi^3}\;\cdot\frac{1}{4\pi} 
         \; J^{Y}.
\label{Mdef3p0}
\end{equation}

\end{itemize}

\noindent
with

\begin{eqnarray}
J^X&=&\frac{m_\pi}{p\,p'}\int_0^\infty dr\,u_{0,1}\,H^X(r)\,u_{1,1}(r)
\\ [0.3cm]
J^Y&=&\frac{m_\pi}{p\,p'}\int_0^\infty dr\,u_{0,0}\,H^Y(r)\,u_{1,0}(r)
\end{eqnarray}

\noindent
where $X$ and $Y$ represents $\Delta$, $IA$, etc.,
$J^X$ and $J^Y$ are dimensionless integrals, 
$H^{X}(r)$ and $H^Y(r)$ are the corresponding operators, 
obtained using the matrix elements given in the Appendix, which are the
same used in the deuteron final state (for the 
$^3P_1\rightarrow$ $^3S_1$ channel) and in the $pp\rightarrow pp\pi^0$
\cite{bira} (for the $^3P_0\rightarrow$ $^1S_0$ channel). 
We will plot $J^{X}$ and $J^Y$ in terms of $p'$ to evaluate the energy 
dependence of the amplitude, 
and $dJ^X/dr$ and $dJ^Y/dr$ in terms of $r$ to study the $r$ dependence of 
the integrand.

For the $^3P_1\rightarrow$ $^3S_1$ channel, the coordinate space 
expressions for the amplitudes are pretty much 
the same of the $S$-wave deuteron amplitudes. For instance, the 
Weinberg-Tomozawa term gives
\begin{equation}
H^{WT}(r)=\frac{3}{4}\;{(1+\tilde{m}_\pi r)\over r^2} e^{-\tilde{m}_\pi r},
\label{wtpn}
\end{equation}
\noindent 
where $\tilde{m}_\pi=\sqrt{\frac{3}{4}}m_\pi$. 
For the $^3P_0\rightarrow ^1S_0$ channel, the expressions follow closely the
work of Cohen {\it et al} \cite{bira}, 
the isospin matrix elements being the only
difference. As we did in the deuteron case, we include the effects of 
form factors when dealing with heavy meson contributions.

The final steps consist of computing the total matrix element ${\cal M}$
for the two channels, using Eq.~(\ref{calm}) for the $^3S_1$ final state 
and the same equation but without WT and the GC terms for the $^1S_0$
final state, since in this state the isovector contributions are zero.

The cross section is obtained by squaring the total amplitude, and 
integrating over the available phase space. We find
\begin{equation}
\sigma=\sum_{spins}\,\frac{1}{v}\int_0^{p_{max}'}\;dp'\;
\frac{{p'}^2\,q}{(2\pi)^3}\;|{\cal M}|^2\;\frac{m_n}{2m_n+w(q)}
\label{crosspn}
\end{equation}
where $v$ is the laboratory velocity of the incident proton, $q$ is 
the pion 3-momentum, $w(q)=\sqrt{q^2+m_\pi^2}$,  
$p$ is the magnitude of the center-of-mass initial 3-momentum and 
${p}_{max}'=\sqrt{p^2-m_N m_\pi}$. The $\Sigma_{spins}$ indicates that 
(a) a sum over final spin states and (b) an average over 
initial spin states must be made, which result in factors of $3/4$
for the $^3S_1$  
and $1/4$ for the $^1S_0$ final states.

\section{Input Parameters and Results}

The various amplitudes considered in the last section depend on
several parameters that we can determine {}from other processes.
The WT, GC to WT, and the impulse-approximation operators depend on 
the pion mass, $m_\pi = 138$ MeV \cite{pdb}, and on   
\begin{equation}
\frac{g_A}{f_{\pi}}=\frac{g_{\pi NN}}{m_N};
\end{equation}
\noindent
we use the value of $g_{\pi NN}$ appropriate for each potential. 
The $\Delta$ operator of Eq.(\ref{Delta}) further depends 
on the $\Delta-N$ mass splitting $\delta = 294$ MeV \cite{pdb} and  
on the $\pi N \Delta$
coupling constant, $h_A$. This has been fixed {}from $P$-wave
$\pi N$ scattering (see, e.g., Ref.\cite{weise}), 
\begin{equation}                             
\frac{h_A}{g_A} \simeq 2.1.
\end{equation}

The seagull operator of Eq.(\ref{ST}) depends on four parameters
$c_{1,2,3}$ and $\delta m_{N}$. 
The $c_{i}$'s can be obtained by fitting $S$-wave $\pi N$ scattering.
In Ref.\cite{BKM95} they were found to be
\begin{equation}
\left[4c_1+\frac{\delta m_N}{4m_\pi^2}-\left(c_2+c_3
-\frac{g_A^2}{8m_N}\right)\right]=-\frac{2.31}{2m_N}\; ,
\end{equation}
\noindent 
from the $\sigma$-term, the isospin-even scattering length, 
and the axial polarizability, to ${\cal O}(Q^3)$.
We refer to this as ``SeaI''.
A different determination from an ${\cal O}(Q^2)$ fit to $\pi N$
sub-threshold
parameters \cite{sub} gives $-0.29/2m_N$ instead.
We refer to this as ``SeaII''.
Newer determinations \cite{ulfandall} give values closer to the more
negative value, but we use both values in order to estimate
the importance of this contribution.
Note that the analysis of Ref.\cite{BKM95} does not include 
the isobar explicitly. 
Since the inclusion of the $\pi N \Delta$ interaction  
only affects $S$ waves at one order higher than the $c_{i}$'s,
the above values can still be used to estimate the effect of
$S$-wave rescattering.  
The parameter, $\delta m_{N}$, can in principle also
be determined {}from S-wave $\pi N$ scattering, but would require
a careful analysis of other isospin-violating effects. Chiral
symmetry relates it to the
strong interaction contribution to the nucleon mass splitting, which
is also difficult to determine directly. 
Estimates of the electromagnetic contribution $\bar{\delta} m_{N}$
are more reliable, $\bar{\delta} m_{N}\sim -1.5$ MeV \cite{jerry},
and give $\delta m_{N} \sim 3$ MeV. To be definite, we use
\begin{equation}
\delta m_{N}=3 \, \mbox{MeV}.
\label{deltmn}
\end{equation}
 
Finally, the $\sigma, \omega$ and the $\rho-\omega$ operators involve 
$g_{h}$, $\Lambda_{h}$, $m_{h}$, and $C_\omega$, parameters listed in
Table A.3 of Ref. \cite{machleidt}. They also involve $g_{\pi\rho\omega}$,
discussed in Ref. \cite{KMR96}.
These heavy-meson contributions 
correspond to chiral Lagrangian coefficients of
natural size:
$d_1^{\prime} + e_1 \approx -1.5 (1/f_{\pi}^2 M)$,  
$e_1 \approx -2 (1/f_{\pi}^2 M)$,
$e_2 \approx 2 (1/f_{\pi}^2 M)$,
and $g \approx 4 (1/f_{\pi}^2 M^3)$.

\subsection{The $pp\rightarrow d\pi^+$ reaction}

\begin{figure}
\vspace{-4.5cm}
\epsfxsize=13.0cm
\centerline{\epsffile{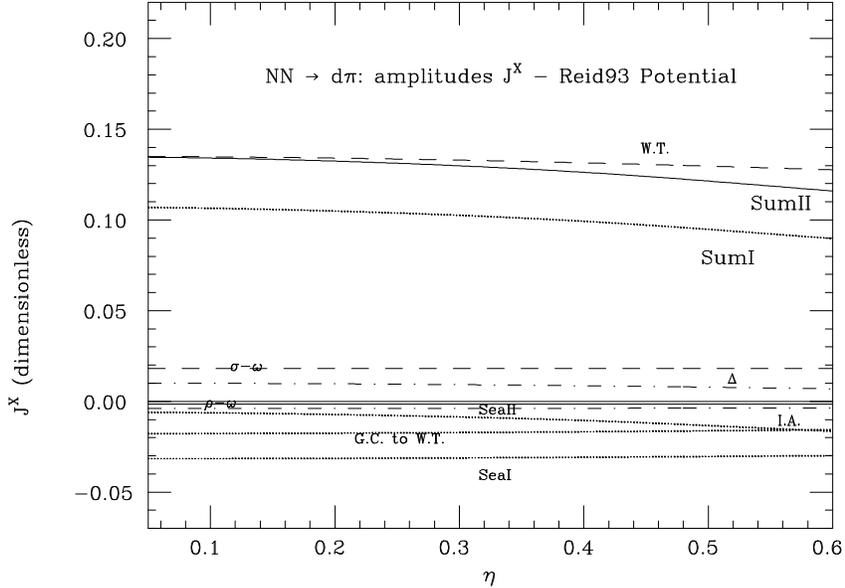}}
\vspace{-4.5cm}
\caption{Matrix elements $J$ as function of $\eta$
for the various contributions to $pp\rightarrow d\pi^+$, 
calculated with wavefunctions
from the Reid93 potential.}
\label{Fig.2}
\end{figure}

\begin{figure}
\vspace{-4.5cm}
\epsfxsize=13.0cm
\centerline{\epsffile{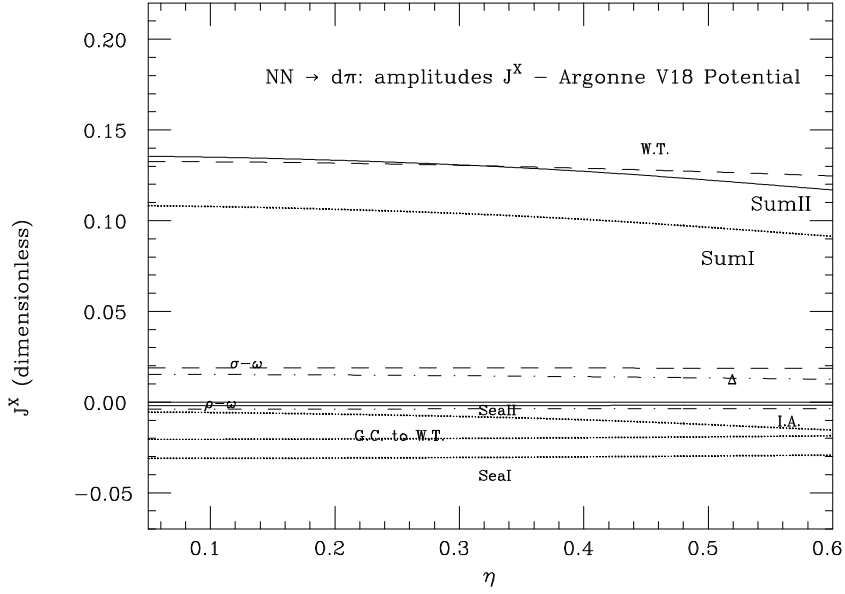}}
\vspace{-4.5cm}
\caption{Matrix elements $J$ as function of $\eta$
for the various contributions to $pp\rightarrow d\pi^+$, 
calculated with wavefunctions
from the Argonne V18 potential.}
\label{Fig.3}
\end{figure}

The relative sizes of the various contributions 
to the matrix element $J$ of this reaction as function of $\eta$
are shown in Fig. \ref{Fig.2} for the Reid93 potential and
in Fig. \ref{Fig.3} for the AV18 potential. There is a very strong similarity
between these two sets of results.

\begin{figure}
\vspace{-4.5cm}
\epsfxsize=13.0cm
\centerline{\epsffile{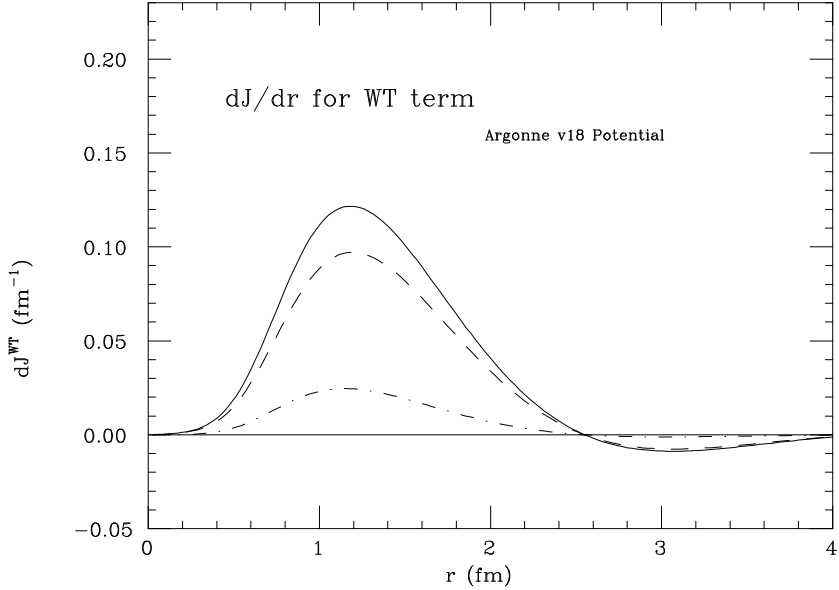}}
\vspace{-4.5cm}
\caption{Typical integrands ($\eta=0.3$) of the Weinberg-Tomozawa
contribution to $pp\rightarrow d\pi^+$ as function of the radial coordinate
$r$ for deuteron $S$ and $D$ waves of the Argonne V18 potential.
The $S$ wave is given by the dashed line, the $D$ wave by the dot-dashed 
line and the sum by the solid line.}
\label{Fig.4}
\end{figure}

\begin{figure}
\vspace{-4.5cm}
\epsfxsize=13.0cm
\centerline{\epsffile{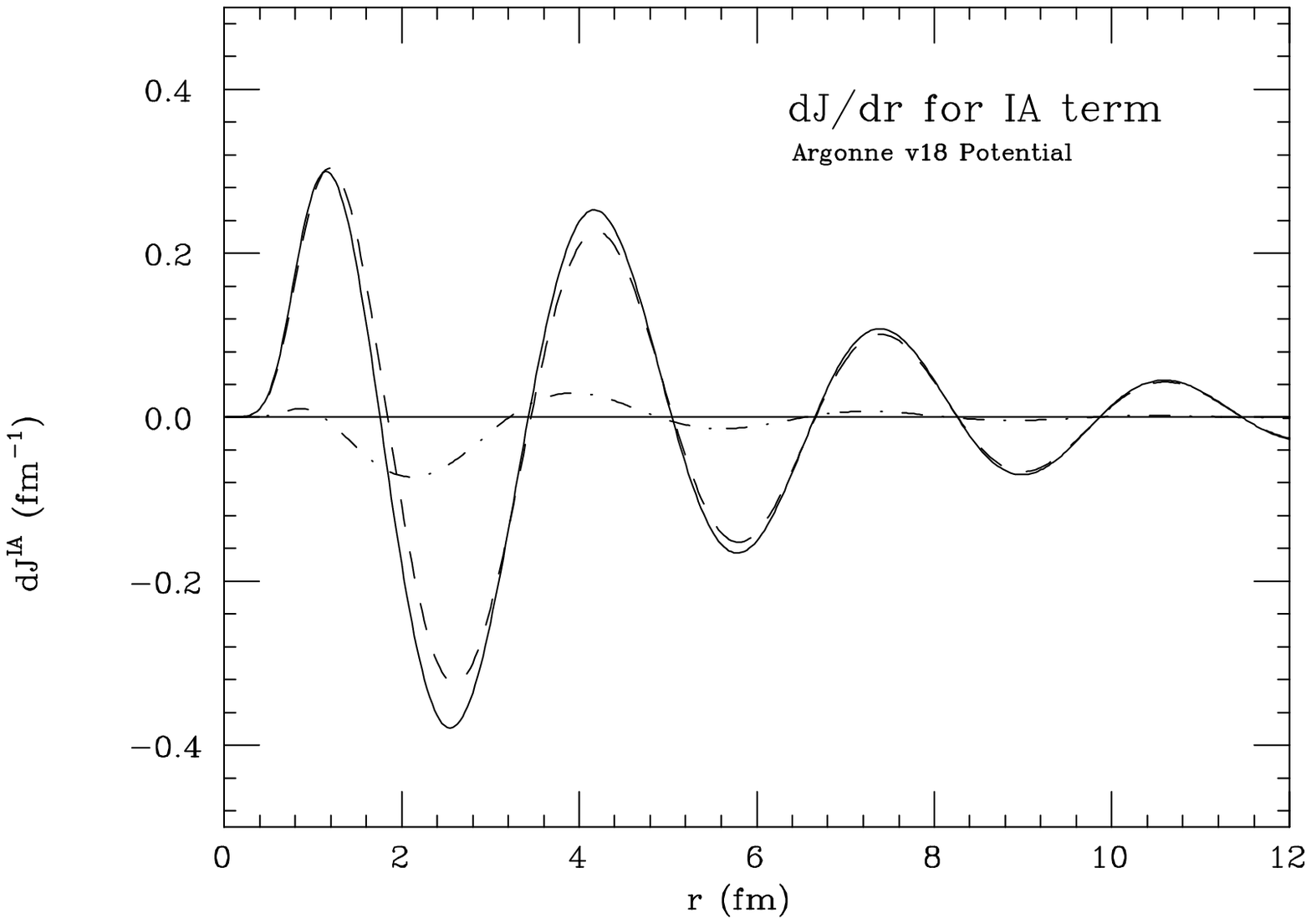}}
\vspace{-4.5cm}
\caption{Typical integrands ($\eta=0.3$) of the impulse
contribution to $pp\rightarrow d\pi^+$ as function of the radial coordinate
$r$ for deuteron $S$ and $D$ waves of the Argonne V18 potential.
Lines are as in Fig. 4.}
\label{Fig.5}
\end{figure}

\begin{figure}
\vspace{-4.5cm}
\epsfxsize=13.0cm
\centerline{\epsffile{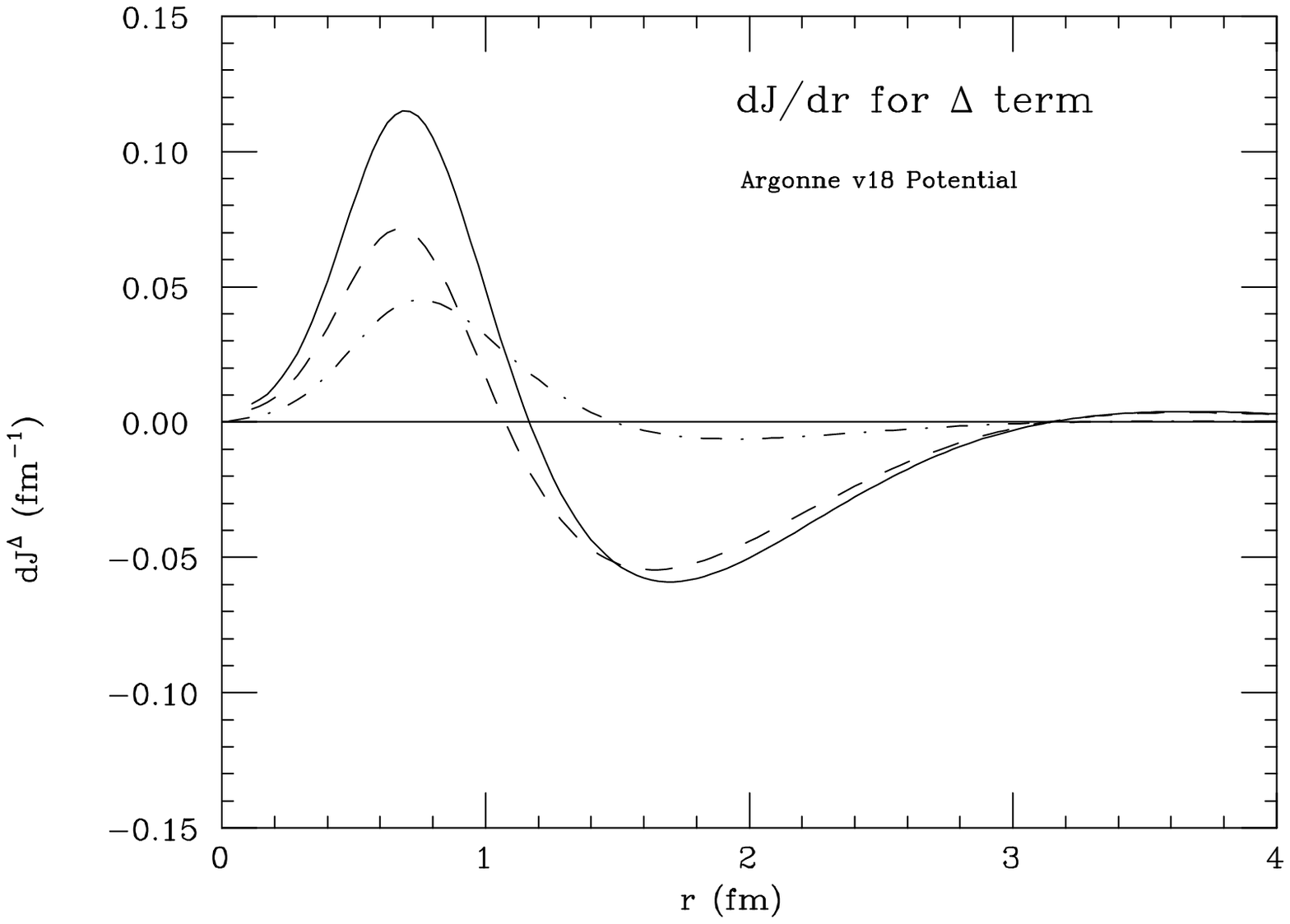}}
\vspace{-4.5cm}
\caption{Typical integrands ($\eta=0.3$) of the delta
contribution to $pp\rightarrow d\pi^+$ as function of the radial coordinate
$r$ for deuteron $S$ and $D$ waves of the Argonne V18 potential.
Lines are as in Fig. 4.}
\label{Fig.6}
\end{figure}

Our leading order comprises WT, IA and $\Delta$. The first
noticeable observation is that WT is by far the largest contribution.
This is a consequence of cancellations in IA and $\Delta$
which were not anticipated by the power counting. 
In Figs.~\ref{Fig.4}, \ref{Fig.5}, and \ref{Fig.6}
we can see typical integrands for the three contributions
at $\eta=0.3$, in the case of the Argonne V18 potential.
While for WT the contributions from the $S$ and $D$ deuteron waves
add and are dominated by the region around $r=1.5$ fm,
for IA and $\Delta$ the $S$ and $D$ waves tend to interfere destructively,
and contributions from different $r$ regions approximately
cancel.

Our formally sub-leading contributions are all
considerably smaller than WT. This suggests that
the theoretical control of this reaction is greater
than for $pp\rightarrow pp\pi^0$.
Moreover, sub-leading contributions come with
different signs, partially cancelling.
In fact, the $\sigma$,$\omega$ and GC 
terms come out with similar size 
---as predicted by the power counting--- but with opposite signs; 
there is an almost complete 
accidental cancellation between them.
$\rho-\omega$ exchange is of higher order and indeed very small.
As a result, when the small contribution from 
the combination of seagull parameters ``SeaII''
is considered, the sum of all sub-leading 
contributions is small. The sum is somewhat larger
and of opposite sign to leading order 
when ``SeaI'' is used.

\begin{figure}
\vspace{-4.5cm}
\epsfxsize=13.0cm
\centerline{\epsffile{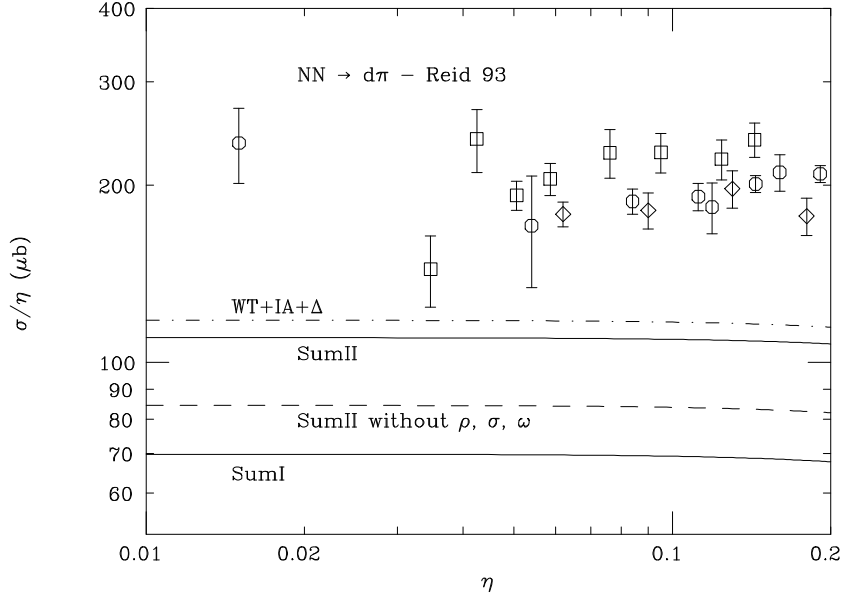}}
\vspace{-4.5cm}
\caption{Reduced cross-section $\sigma/\eta$ 
for $pp\rightarrow d\pi^+$ as function of $\eta$
calculated with the Reid93 potential
in leading order ($WT+IA+\Delta$), in leading plus sub-leading order 
with two sets of parameters (SumI and SumII),
and in leading plus sub-leading order 
without heavy-meson exchange (SumII without $\rho$, $\sigma$, $\omega$), and
compared with data from TRIUMF (circles) \protect\cite{triumf}, 
COSY (diamonds) \protect\cite{cosy}  and IUCF (squares) \protect\cite{iucf}.}
\label{Fig.10}
\end{figure}

\begin{figure}
\vspace{-4.5cm}
\epsfxsize=13.0cm
\centerline{\epsffile{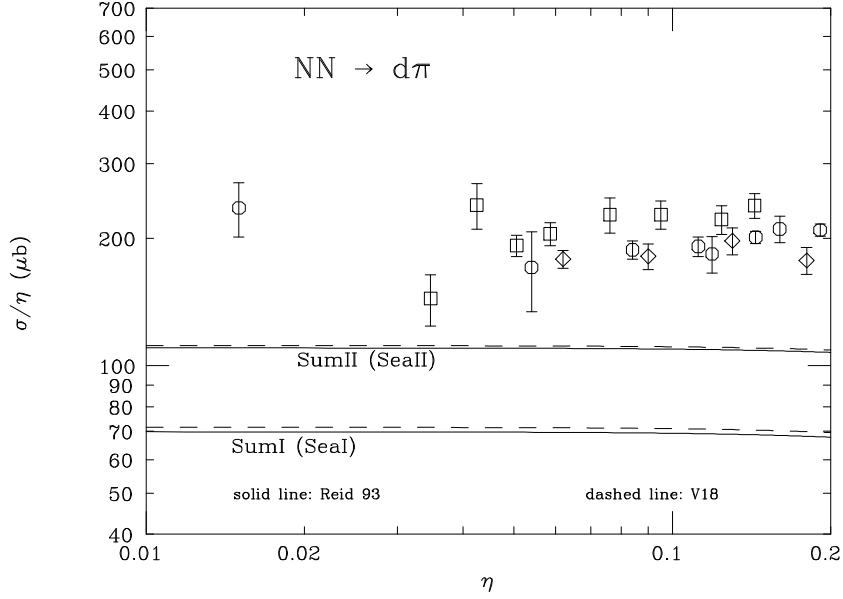}}
\vspace{-4.5cm}
\caption{Reduced cross-section $\sigma/\eta$ for $pp\rightarrow d\pi^+$
as function of $\eta$
for the sum of all the contributions considered
with two sets parameters (SumI, SumII) for
the Reid93 (solid line) and Argonne V18 (dashed line) potentials,
compared with data from TRIUMF (circles) \protect\cite{triumf}, 
COSY (diamonds) \protect\cite{cosy}  and IUCF (squares) \protect\cite{iucf}.}
\label{Fig.11}
\end{figure}

In Fig. \ref{Fig.10} we compare our 
leading  and sub-leading results for the $pp$ cross-section
(without Coulomb) using the Reid93 potential 
with data from Refs. \cite{triumf,cosy,iucf}.
We see that our curves are approximately constant, as expected.
Recent data points for $\eta<0.1$ are not all consistent, but they cluster
around an average $\alpha$ of about 180 to 190 $\mu$b. 
The value of $\alpha$ in leading order (WT+IA+$\Delta$)
is a factor of about 1.5 below data.
The destructive interference with next-order contributions
increases the disagreement by an amount depending
on the value of $\pi N$ isoscalar rescattering term.
The change is bigger when ``SeaI'' is used. This can be
seen in Fig. \ref{Fig.10} as ``SumI'' and ``SumII''.
Because of the cancellations among sub-leading terms,
the result exhibits comparable dependence
on the short-range contributions, 
as can be seen in Fig. \ref{Fig.10}
as ``SumII without $\sigma$, $\rho$, $\omega$''. 
We summarize our results in Fig. \ref{Fig.11} where we show
the sum of all the contributions we considered
for the two sets of seagull parameters, and for the two
potentials. 

\subsection{The $pp\rightarrow pn\pi^+$ reaction}

\begin{figure}
\vspace{-4.5cm}
\epsfxsize=13.0cm
\centerline{\epsffile{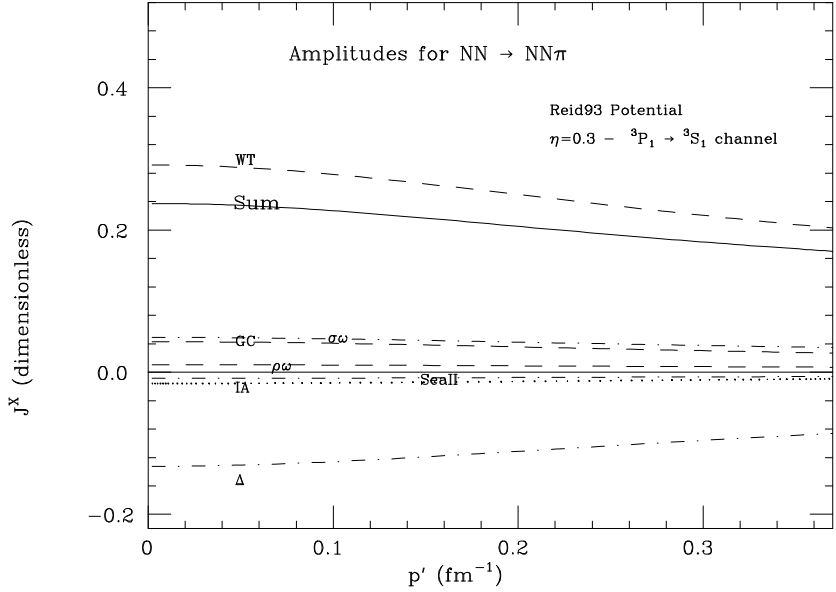}}
\vspace{-4.5cm}
\caption{Matrix elements $J$ as function of $p'$
for the various contributions to $pp\rightarrow pn\pi^+$, 
calculated with wavefunctions
from the Reid93 potential: $^3S_1$ final state.}
\label{Fig.12}
\end{figure}

\begin{figure}
\vspace{-4.5cm}
\epsfxsize=13.0cm
\centerline{\epsffile{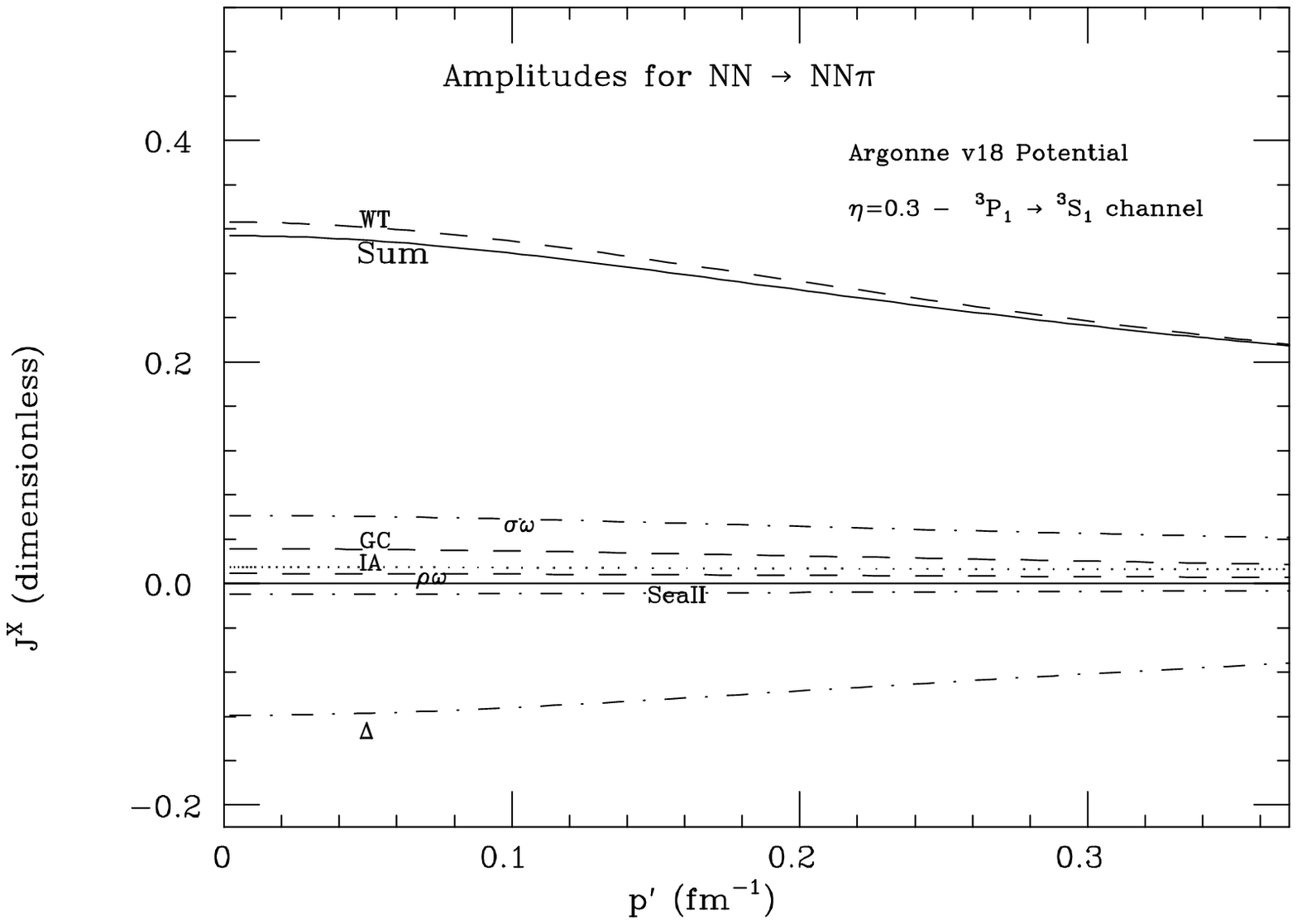}}
\vspace{-4.5cm}
\caption{Matrix elements $J$ as function of $p'$
for the various contributions to $pp\rightarrow pn\pi^+$, 
calculated with wavefunctions
from the Argonne V18 potential: $^3S_1$ final state.}
\label{Fig.13}
\end{figure}

\begin{figure}
\vspace{-4.5cm}
\epsfxsize=13.0cm
\centerline{\epsffile{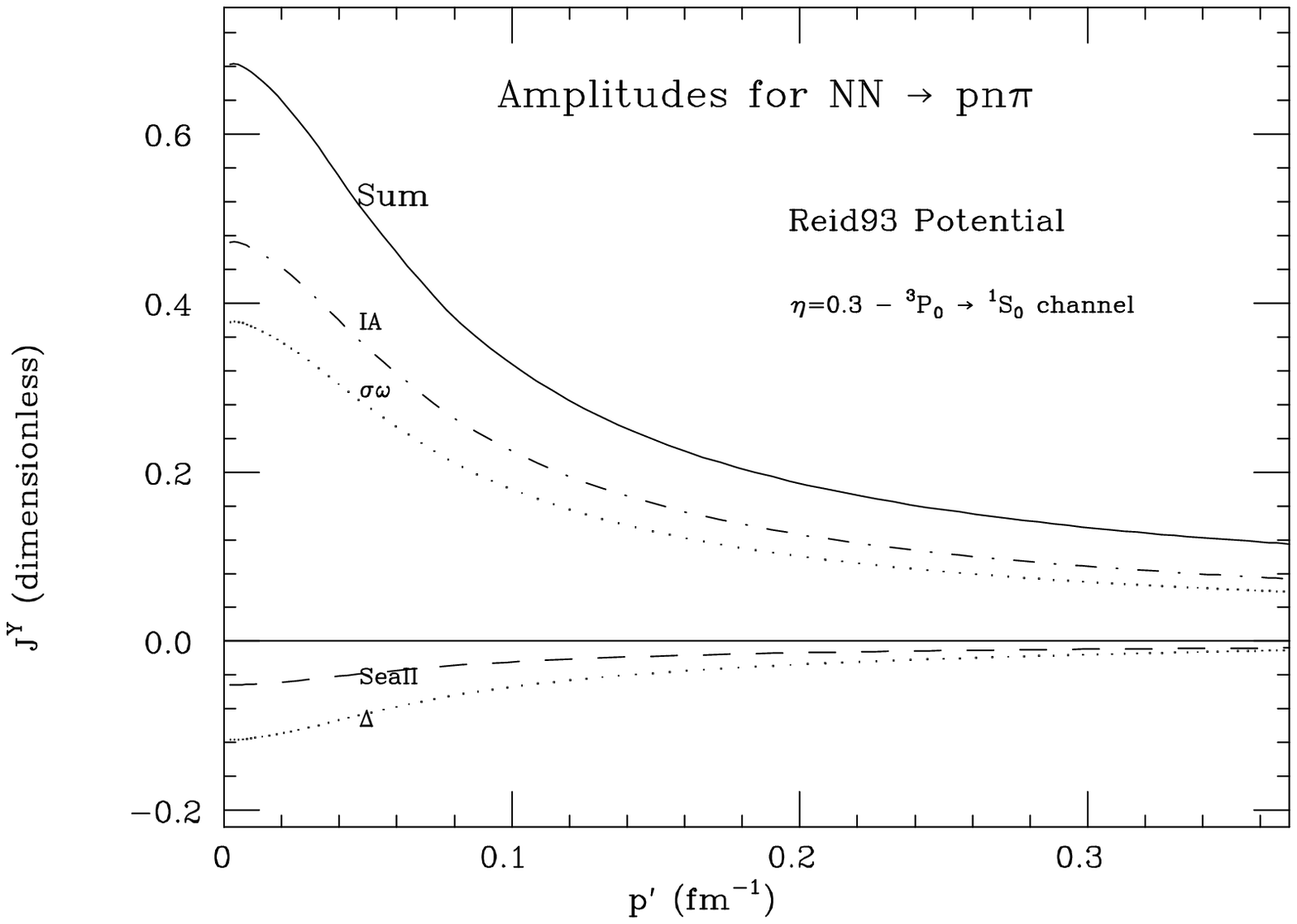}}
\vspace{-4.5cm}
\caption{Matrix elements $J$ as function of $p'$
for the various contributions to $pp\rightarrow pn\pi^+$, 
calculated with wavefunctions
from the Reid93 potential: $^1S_0$ final state.}
\label{Fig.14}
\end{figure}

\begin{figure}
\vspace{-4.5cm}
\epsfxsize=13.0cm
\centerline{\epsffile{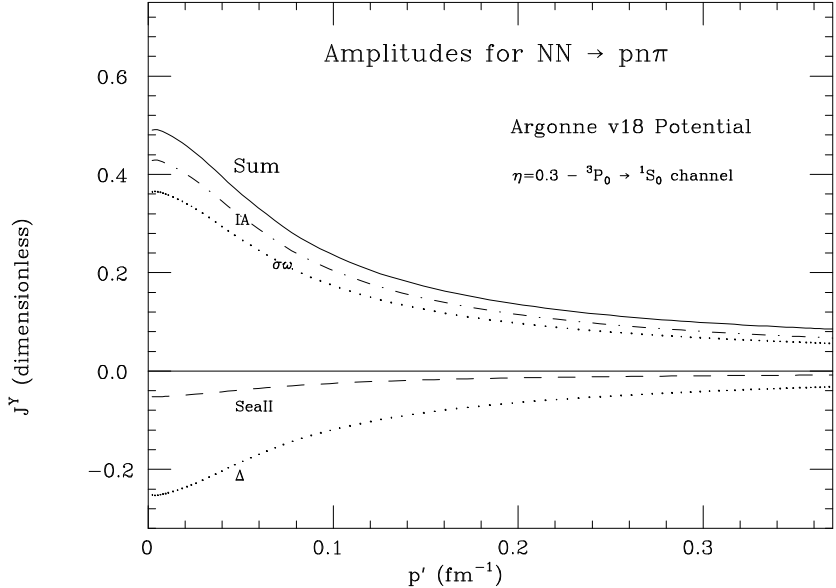}}
\vspace{-4.5cm}
\caption{Matrix elements $J$ as function of $p'$
for the various contributions to $pp\rightarrow pn\pi^+$, 
calculated with wavefunctions
from the Argonne V18 potential: $^1S_0$ final state.}
\label{Fig.15}
\end{figure}

The relative sizes of the various contributions
to the matrix element $J$ of this reaction as function of $p'$
are  shown for the $^3S_1$ final state
in Fig. \ref{Fig.12} for the Reid93 potential and
in Fig. \ref{Fig.13} for the AV18 potential;
and for the $^1S_0$ final state
in Fig. \ref{Fig.14} for the Reid93 potential and
in Fig. \ref{Fig.15} for the AV18 potential.

Here again the WT contribution is the largest;
since it contributes only to the $^3S_1$ final state,
this channel is dominant.
Most of the other contributions are much smaller and tend to
cancel to some extent. 
The exception is the $\Delta$,
which has a significant destructive interference with WT
in the $^3S_1$  state.
The IA contribution is small due to the same type of
cancellation observed before among different regions in coordinate
space.

\begin{figure}
\vspace{-4.5cm}
\epsfxsize=13.0cm
\centerline{\epsffile{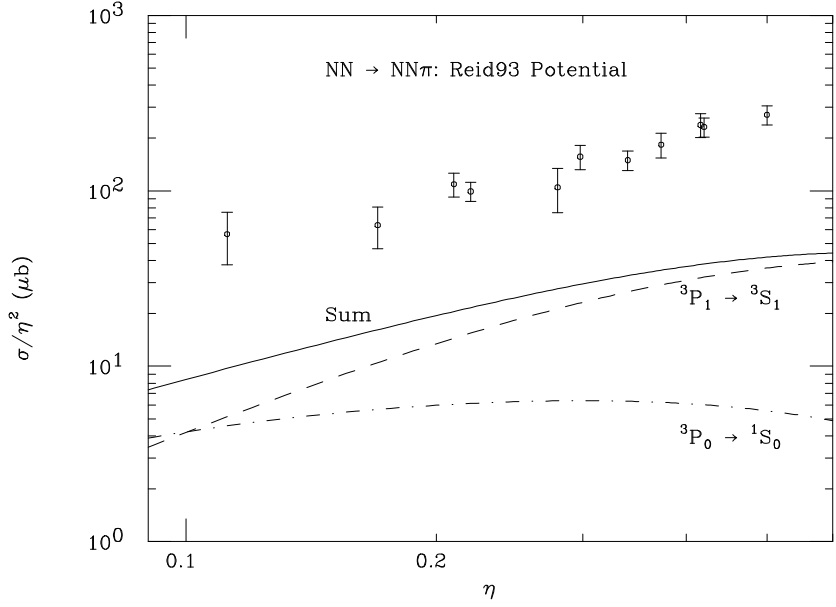}}
\vspace{-4.5cm}
\caption{Reduced cross-section $\sigma/\eta^2$ 
for $pp\rightarrow pn\pi^+$ as function of $\eta$
calculated with the Reid93 potential:
$^3S_1$ final state (dashed line), 
$^1S_0$ final state (dash-dotted line),
their sum (solid line),
and  data from IUCF \protect\cite{steve1,steve2}.}
\label{Fig.19}
\end{figure}

\begin{figure}
\vspace{-4.5cm}
\epsfxsize=13.0cm
\centerline{\epsffile{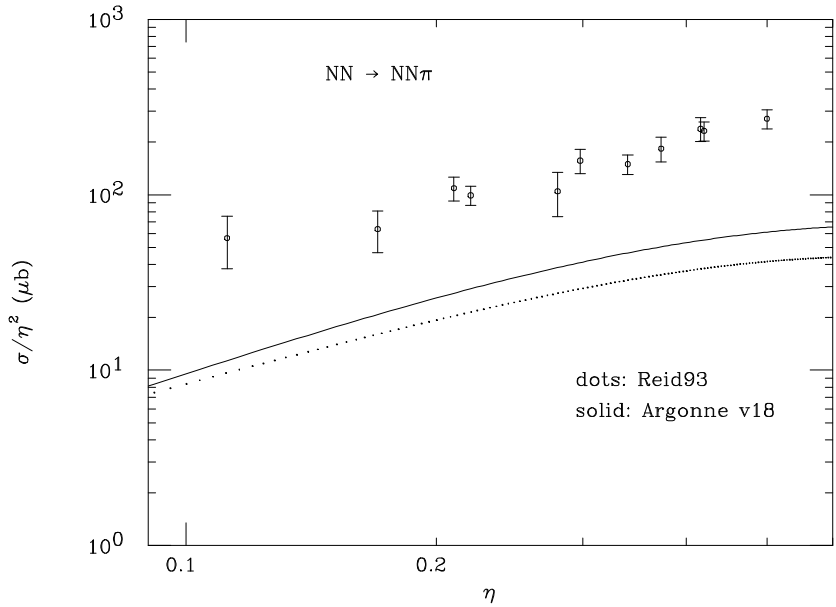}}
\vspace{-4.5cm}
\caption{Reduced cross-section $\sigma/\eta^2$ 
for $pp\rightarrow pn\pi^+$ as function of $\eta$
for the sum of all the contributions: 
Reid93 potential (dotted line),
Argonne V18 potential (solid line),
and  data from IUCF \protect\cite{steve1,steve2}.}
\label{Fig.20}
\end{figure}

In Fig. \ref{Fig.19} we compare our 
results for the different final states
using the Reid93 potential 
with data from Ref. \cite{steve1,steve2}.
In Fig. \ref{Fig.20} we summarize our results for
the two potentials considered and compare them to the same data.
We see that the theory produces a correct shape for the 
$\eta$ dependence but fails in magnitude by a factor of $\sim 5$.
Use of ``SeaI'' further worsens the results. 
Once again the differences between the two potentials are minimal.

\subsection{Discussion}

The $d\pi^+$ final state has been considered before in the literature.
Ref. \cite{KR66} has found that the impulse term was very small due to
the strong cancellation between the $S$ and $D$ waves in the matrix element
shown above. Mostly due to the WT,
$\alpha$ of 146 to 160 $\mu$b was
obtained  
using the older, higher value of the $\pi NN$ coupling constant;
using the more recent value,
we get 124 to 136 $\mu$b.
A similar analysis \cite{horowitz}
included also 
form factors at the $\pi\pi NN$ vertices, which led to a decrease 
of approximately 20\% in the cross section; with the more modern
coupling constants, the overall result was near 100 $\mu$b.
Recently, Ref. \cite{hanhart} 
re-analyzed this reaction using a covariant
approach. The result for $\alpha$ using only the WT term was 
again near 100 $\mu$b, but the amplitude contains what we refer to as the
Galilean correction to the WT term, 
which has an opposite sign. 
Our results for the WT term alone are in numerical
agreement with these works, since we find that it gives
an $\alpha$ of about 100 $\mu$b.

Other contributions tend to worsen the description of the data.
Particularly damaging are the contributions of the rescattering type.
Here, as for $pp\rightarrow pp\pi^0$, the $S$-wave seagulls
tend to interfere destructively with the leading mechanism.

\begin{figure}
\vspace{-4.5cm}
\epsfxsize=13.0cm
\centerline{\epsffile{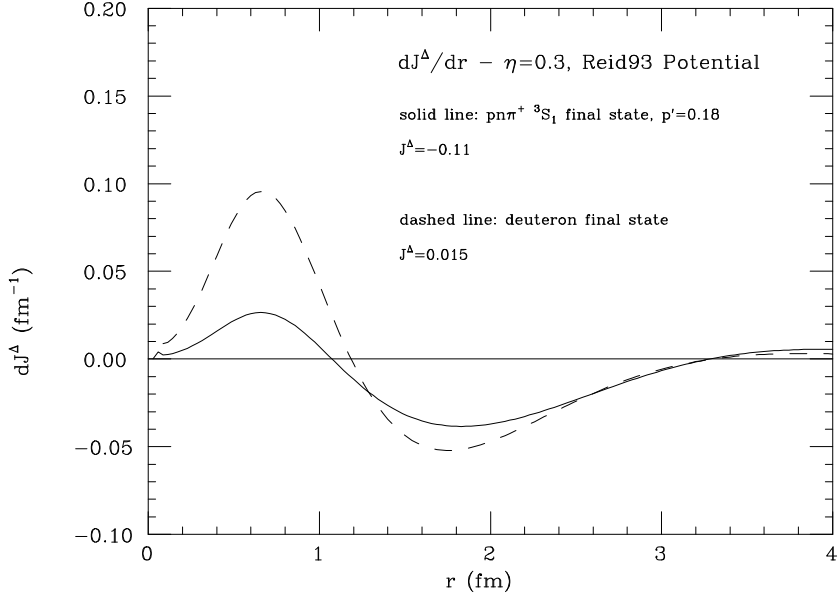}}
\vspace{-4.5cm}
\caption{Typical integrands ($\eta=0.3$) of the delta
contribution as function of the radial coordinate
$r$ for the $^3S_1$ $pn$ final state at $p'=0.18$ (solid line)
compared to the deuteron final state (dashed line).}
\label{Fig.21}
\end{figure}

As in the case of the $pp\rightarrow pp\pi^0$ reaction,
the $\Delta$ contribution shows appreciable dependence
on the potential used. But unlike that reaction, here
it generates relatively small dependence on the potential in the final result,
as consequence of the large relative size of WT.
The $\Delta$ contribution is the one that presents
the largest change when we compare $pn$ and $d$ final states.
In Fig. \ref{Fig.21} we compare the radial distributions
of the delta contribution for the $^3S_1$ $pn$ final state at 
$p'=0.18$ fm$^{-1}$ 
and for the deuteron final state (same as in Fig. \ref{Fig.6}).
{}From this we can see how this contribution has different
signs in the two channels, being small and constructive with
WT for $d$ ($J^{WT}_d=0.130$ and $J^{\Delta}_d=0.015$; $\eta=0.3$)
and larger and destructive for $pn$
($J^{WT}_{pn}=0.260$ and $J^{\Delta}_{pn}=-0.11$; $\eta=0.3$; $p'=0.18$)

Note, however, that the $\Delta$ contribution is subject to large
uncertainties.
First, the $\pi N\Delta$ coupling constant is not well determined
and appears squared; although we do not know whether this coupling should
be bigger or smaller than the value used here, it could
contribute to a decrease of the $\Delta$ effect.
Second, the $\pi N\Delta$ form factor seems to be much
softer than the corresponding $\pi NN$ form factor;
our neglecting both enhances the $\Delta$ contribution
relative to the WT term.
Third, we have neglected the $\Delta-N$ mass difference
in energies; since the $\Delta$ mass would appear in the denominator,
the delta contribution would decrease by a factor of 
$\sim 2m_N/(m_N+m_\Delta)$.
Fourth, we have neglected the kinetic energy 
of the delta, account of which would further decrease
its amplitude.
Fifth, it is known that there can be some cancellation
between the pion- and rho-exchange delta terms; including
the latter would also diminish the delta contribution.
Assuming each of these gives a 10\% decrease, we could very well be
overestimating the  $\Delta$ contribution by 50\% or more.

\begin{figure}
\vspace{-4.5cm}
\epsfxsize=13.0cm
\centerline{\epsffile{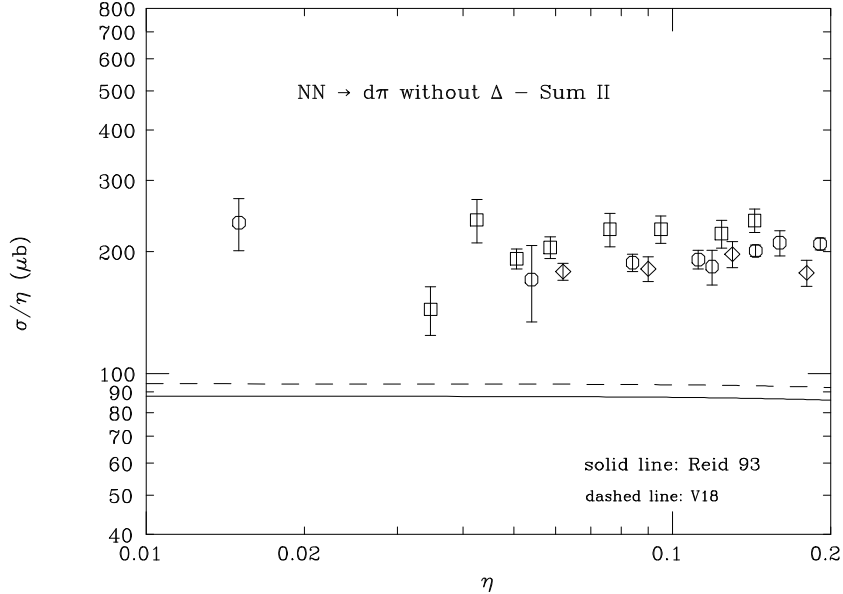}}
\vspace{-4.5cm}
\caption{Reduced cross-section $\sigma/\eta$ for $pp\rightarrow d\pi^+$
as function of $\eta$. The graph show the sum of all the contributions, 
excluding the $\Delta$: Reid93 potential (dotted line),
Argonne V18 potential (solid line),
and data from TRIUMF (circles) \protect\cite{triumf}, 
COSY (diamonds) \protect\cite{cosy}  and IUCF (squares) 
\protect\cite{iucf}.}
\label{Fig.22}
\end{figure}

\begin{figure}
\vspace{-4.5cm}
\epsfxsize=13.0cm
\centerline{\epsffile{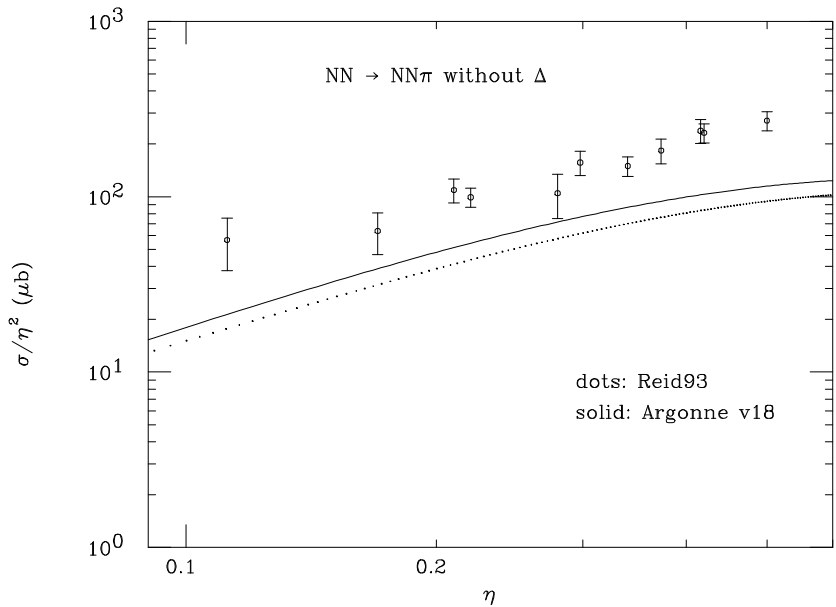}}
\vspace{-4.5cm}
\caption{Reduced cross-section $\sigma/\eta^2$ for $pp\rightarrow pn\pi^+$ 
as function of $\eta$. The lines shown the
sum of all the contributions, excluding the $\Delta$: 
Reid93 potential (dotted line),
Argonne V18 potential (solid line),
and  data from IUCF \protect\cite{steve1,steve2}.}
\label{Fig.23}
\end{figure}

If we neglect this contribution altogether, 
we obtain the results in 
Fig. \ref{Fig.22} for the cross section in the $d$ channel
and in Fig. \ref{Fig.23} for the cross section in the $pn$ channel.
This brings theory to underestimate both sets of data
by a common factor of $\simeq 2$.
This implies that there must be further corrections to the amplitude of
about 50\%.
This is not unlike the $pp\rightarrow pp\pi^0$ reaction considered 
in Ref. \cite{KMR96},
where theory tends to fail by a similar factor.
The case for failure of theory there is less clear-cut, however,
because of the lack of a large contribution as for the WT
term here. As a consequence, the usually small effect
of other mechanisms is enhanced and the result is dominated
by shorter-range dynamics; more sensitivity
to the potential and seagull terms surfaces, 
and it is possible to find a combination of parameters that
includes data \cite{KMR96}.
No such gimmicks work here.
For example, we find that
heavy-meson exchange ---hailed as solution in the
$pp\rightarrow pp\pi^0$ reaction---
does not help much in $\rightarrow d\pi^+$ and $\rightarrow pn\pi^+$,
in agreement with the findings in Refs. \cite{niskanen,unpharry}.
The dependence on the $\Delta$ contribution (which is a particular
type of rescattering) suggests that we need better
control over longer-range contributions, such as $\pi N$ rescattering
and two-pion exchange.

\section{Conclusion}

We have calculated the cross section near threshold for the
reactions $pp \rightarrow d \pi^+, \rightarrow pn \pi^+$,
using a chiral power counting to order interactions.
The interactions included contain most of the interactions
found in the literature. In particular, they include
mechanisms used in various models of the
$pp\rightarrow pp\pi^0$ reaction.
In contrast to the latter,
results here depend relatively little on the wavefunction
employed and on short-range interactions.
The Weinberg-Tomozawa term dominates, but the delta contribution
can be important, with large uncertainty. 
Our computed results typically
fall a factor of about two below the data. Our kernels vary as the
cube of a generic meson-nucleon coupling constant, so such discrepancy
can be parametrized as a 12\% deficiency in the coupling constants. The
pion-nucleon
coupling constant is known to higher precision than that, but not to much
higher precision. 
  Thus the discrepancy we find might not be a very serious problem.
  
Our main conclusion 
is that a relatively long-range mechanism ---
such as $\pi N$ rescattering and/or two-pion exchange--- is needed
for the description of these reactions.
Accordingly, we suggest that advance in understanding pion production
in $NN$ collisions must follow
not from the study of $pp\rightarrow pp\pi^0$ by itself
---as has been the trend of theoretical study to date---
but from focus on an understanding of long-range effects
that afflict all channels.

\vspace{2cm}
{\large \bf Acknowledgments} 

We thank C. Hanhart for a helpful discussion.
This research was supported in part by the U.S. DOE
under grant DE-FG03-97Er41014 and NSF under grant PHY 94-20470.
The work of C.A.dR. was supported by the Brazilian FAPESP 
under contract numbers 97/05817-0 and 97/6209-4. He thanks
the Nuclear Theory group at the University of Washington
for hospitality during the initial stages of this work.

\vspace{1.5cm}
{\Large \bf Appendix}
\vspace{.5cm}

The calculation of the cross-section in Section IV relies
on the following matrix elements.

For isospin, 

\begin{eqnarray}
&&\langle 00\,|\;i\epsilon_{abc}\;\tau_b^{(1)}\tau_c^{(2)}\,|11\rangle =
-2 \;,\\
&&\langle 00\,|\vec{A}\,\tau_a^{(1)}-\vec{B}\,\tau_a^{(2)}\,|11\rangle =
\vec{A}+\vec{B} \;,\\
&&\langle 00\,|\vec{A}\,\tau_3^{(1)}\tau_a^{(2)}-\vec{B}\,\tau_a^{(1)}
\tau_3^{(2)}\,|11\rangle = -\left(\vec{A}+\vec{B}\right) \;,\\
&&\langle 00\,|\,\delta_{3a}\,\vec\tau^{(1)}\cdot\vec\tau^{(2)}\,|11\rangle=0 \;, \\ 
&&\langle 00\,|\frac{1}{2}\left[\tau^{(1)}_a + \tau^{(2)}_a\right]\,|11\rangle=0\;,
\end{eqnarray}

\noindent 
where $\vec{A}$ and $\vec{B}$ are spin operators or scalar products.
The index $a$ represents the isospin of the emerging pion, 
$\langle 00\,|$ is the isospin 
state of the deuteron and $|11\rangle$ the $pp$ isospin initial state.

For spin,
\begin{eqnarray}
&&\langle ^3S_1|\vec{S}\cdot \hat r |^3P_1\rangle = -\sqrt{\frac{2}{3}}\;, \\
&&\langle ^3D_1|\vec{S}\cdot \hat r |^3P_1\rangle = -\sqrt{\frac{1}{3}}\;, \\
&&\langle ^3S_1|i \vec{S}\cdot\vec{p}\,|^3P_1\rangle = - 
\sqrt{\frac{2}{3}}\left({\partial\over\partial r}+{2\over r}\right)\; ,\\
&&\langle ^3D_1|i \vec{S}\cdot\vec{p}\,|^3P_1\rangle = - 
\sqrt{\frac{1}{3}}\left({\partial\over\partial r}-{1\over r}\right)\; ,\\
&&\langle ^3S_1|\vec{\sigma}^{(1)}\cdot\vec{\sigma}^{(2)}\; i \vec{S}\cdot\vec{p}\,|^3P_1\rangle = - 
\sqrt{\frac{2}{3}}\left({\partial\over\partial r}+{2\over r}\right)\; ,\\
&&\langle ^3D_1|\vec{\sigma}^{(1)}\cdot\vec{\sigma}^{(2)}\;i \vec{S}\cdot\vec{p}\,|^3P_1\rangle = - 
\sqrt{\frac{1}{3}}\left({\partial\over\partial r}-{1\over r}\right)\; ,\\
&&\langle ^3S_1|\hat{S}_{12}\,i \vec{S}\cdot\vec{p}\,|^3P_1\rangle = - 2
\sqrt{\frac{2}{3}}\left({\partial\over\partial r}-{1\over r}\right)\; ,\\
&&\langle ^3D_1|\hat{S}_{12}\,i \vec{S}\cdot\vec{p}\,|^3P_1\rangle = - 2
\sqrt{\frac{1}{3}}\left({\partial\over\partial r}+{5\over r}\right)\; .
\end{eqnarray}

\noindent 
where $\hat{S}_{12}$ is the usual tensor operator.

\end{document}